\documentclass[11pt]{article}
\usepackage{eurosym}
\usepackage{amssymb}
\usepackage{graphicx}
\usepackage{amsmath}
\usepackage{makeidx}
\usepackage{indentfirst}

\newcounter{resultnum}[section]
\newcounter{conclusionnum}[section]
\newcounter{conditionnum}[section]
\newcounter{conjecturenum}[section]
\newcounter{examplenum}[section]
\newcounter{exercisenum}[section]
\newcounter{lemmanum}[section]
\newcounter{notationnum}[section]
\newcounter{theoremnum}[section]
\newcounter{definitionnum}[section]
\newcounter{corollarynum}[section]
\newcounter{remarknum}[section]
\newcounter{propositionnum}[section]
\newcounter{acknowledgementnum}[section]
\newcounter{algorithmnum}[section]
\newcounter{axiomnum}[section]
\newcounter{casenum}[section]
\newcounter{claimnum}[section]
\newcounter{summarynum}[section]
\newcounter{problemnum}[section]
\begin{document}

\title{Principles of Einstein--Finsler Gravity and Perspectives in Modern
Cosmology}
\date{July 29, 2012}
\author{\textbf{Sergiu I. Vacaru} \thanks{%
sergiu.vacaru@uaic.ro, Sergiu.Vacaru@gmail.com} \\
\textsl{\small University "Al. I. Cuza" Ia\c si, Science Department,} \\
\textsl{\small 54 Lascar Catargi street, Ia\c si, Romania, 700107 } }
\maketitle

\begin{abstract}
We study the geometric and physical foundations of Finsler gravity theories
with metric compatible connections defined on tangent bundles, or (pseudo)
Riemannian manifolds, endowed with nonholonomic frame structure. There are
considered several generalizations and alternatives to Einstein gravity
including modifications with broken local Lorentz invariance. It is also
shown how such theories (and general relativity) can be equivalently
re--formulated in Finsler like variables. We focus on prospects in modern
cosmology and Finsler acceleration of Universe. Einstein--Finsler gravity
theories are elaborated following almost the same principles as in the
general relativity theory but extended to Finsler metrics and connections.
Finally, some examples of generic off--diagonal metrics and generalized
connections, defining anisotropic cosmological Einstein--Finsler spaces are
analyzed; certain criteria for the Finsler accelerating evolution are
formulated.
\end{abstract}


\section{Introduction}

During last 30 years, the experimental data and existing methodology and
phenomenology of particle physics, and gravity, imposed an interpretation
doctrine that models of Finsler like spacetimes (with metrics and
connections depending on ''velocity/momenta'') are subjected to strong
experimental restrictions. Such theories were not included in the standard
paradigm of modern physics (see respective arguments in Refs. \cite%
{beken,will}\footnote{%
those studies did not include all fundamental geometric/ physical objects in
Finsler geometry/ gravity, for instance, the nonholonomic structure,
nonlinear connections, N--connections, and new types of linear connections
which are adapted to N--connections, the possibility to model (pseudo)
Finsler geometric models as exact solutions in Einstein gravity etc}).

Nevertheless, there are various theoretical arguments \cite%
{mavromatos,mavromatos1,lammer,vcrit,girelli,sindoni,visser} that quantum
gravity models positively result in nonlinear dispersion relations depending
on velocities/momenta. Anisotropic quasi--classical Finsler configurations
originating from quantum gravity are not obligatory restricted for some
inflationary cosmological models and may have important contributions to
dark energy and dark matter in Universe. This constrains us to investigate
Finsler type spacetimes both in quantum gravity theories and modern
cosmology.

In a survey \cite{vrflg} oriented to non--experts in Finsler geometry (but
researches in particle physics and gravity) we discussed in details and
formulated well defined criteria how the Finsler geometry methods and
theories with nonholonomic distributions\footnote{\label{fnnhm}A pair $(%
\mathbf{V},\mathcal{N})$, where $\mathbf{V}$ is a manifold and $\mathcal{N}$
is a nonintegrable distribution on $\mathbf{V}$, is called a nonholonomic
manifold. Modeling Finsler like geometries in Einstein gravity, we have to
consider that $\mathbf{V}$ is a four dimensional (pseudo) Riemannian
spacetime when the Levi--Civita connection is correspondingly deformed to a
linear connection adapted to a N--connection structure defined by a
nonintegrable $(2+2)$--splitting (i.e. nonholonomic distribution for frame
fields). The boldface symbols will be used for nonholonomic manifolds/
bundles and geometric objects on such spaces, as we discuss in details \cite%
{vrflg,ijgmmp,vsgg}.} can be elaborated following the standard paradigm of
modern physics. The purpose of the present work is to study the fundamental
principles of Einstein--Finsler gravity theories\footnote{%
such a model of gravity is constructed following the same principles as the
general relativity theory but on tangent bundles, or on (pseudo) Riemannian
manifolds endowed with nonholonomic distributions, for a (Finsler type)
metric compatible linear connection adapted to a nonlinear connection
structure} and analyze possible further applications in modern cosmology.

Almost all classical gravitational effects are described in the framework of
General Relativity (GR). Recently, it was proposed that certain exceptions
in theoretical cosmology may be related to the dark matter and dark energy
problems and some approaches were formulated for ''nonmetric'' Finlser
gravity models. Here we note that the existence of anisotropies and
inhomogeneities has become a conventional feature in cosmology physics. A
number of cosmological models with Finsler metrics have been elaborated by
now, and this number seems to grow rapidly \cite%
{stavr1,stavr2,lichang5,lichang6}. It is also expected that small
corrections with violations of equivalence principle and local Lorentz
invariance have to be considered in low energy limits within general
approaches to quantum gravity, see a series of works related to Finsler
geometry \cite{mavromatos,girelli,lammer,sindoni,visser,vqgrbr} and
references therein.

There are two general classes of Finsler type gravity theories with very
different implications in physics, mechanics and cosmology. The first class
originates from E. Cartan works on Finsler geometry \cite{cartan}, see
further geometrical developments and applications in \cite%
{horvath,matsumoto,ma,vstr2,vncg,vsgg,vrflg}. In those works a number of
geometric and physical constructions and Finsler geometry methods were
considered for both types of metric compatible or noncompatible connections.
The most related to \textquotedblright standard physics\textquotedblright\
constructions were elaborated for the metric compatible Cartan and canonical
distinguished connections (in brief, d--connections, see details in \cite%
{ma,vrflg}) following geometric and physical principles which are very
similar to those used for building the general relativity theory \cite{vcrit}%
. In the second class of theories, there are Finsler geometry and gravity
models derived for the Berwald and Chern d--connections which are not metric
compatible, see details in \cite{bcs,lichang5,lichang6}; summaries of
results and applications to (non) standard physical theories are given in
Part I of \cite{vsgg} and \cite{vrflg}.

The article is organized as follows: In section \ref{spfls}, we provide
physical motivations for Finsler gravity theories and explain in brief how
fundamental Finsler geometric objects are defined of tangent bundles and in
Einstein gravity. In section \ref{sfeftg}, we consider the gravitational
field equations for Finsler--Einstein gravity. We briefly discuss how
generic off--diagonal\footnote{%
which can not be diagonalized by coordinate transforms} solutions can be
constructed in exact form and provide some examples. Two classes of
cosmology diagonal and off--diagonal solutions on tangent bundles modeling
Finsler acceleration of Universe are analyzed in section \ref{smfctb}.
Finally, in section \ref{concl} we outline the approach and formulate
conclusions. In Appendices, we provide local formulas and examples of
cosmological solutions.

\section{(Pseudo) Finsler Spacetimes}

\label{spfls}In this section, the most important geometric constructions and
physical motivations for Finsler gravity are summarized. We shall follow the
system of notations proposed in \cite{vrflg,vsgg} (see details and
references therein\footnote{%
in this work, we do not aim to provide an exhaustive bibliography (it is not
possible to list and discuss tenths of monographs and thousands of articles
on Finsler geometry and generalizations and various extensions of the
general relativity theory}).

\subsection{Physical motivations for Finsler gravity theories}

\label{ssvls}The first example of a Finsler metric \cite{finsler} was given
by B. Riemann \cite{riemann} who considered forth order forms instead of
quadratic line elements, see historical remarks in \cite{ma,bcs} and, in
relation to standard and non--standard physical theories, in \cite%
{vrflg,vsgg,ijgmmp,vncg}. In this subsection, we show that Finsler like
nonlinear line elements are generated naturally by deformations of standard
Minkowski metrics in special relativity, SR (see also \cite%
{amelino,magueijo,kimberly,ellis,mavromatos1,mavromatos2,stavr1,lammer}).

\subsubsection{Violations of Lorentz symmetry and Finslerian Hessians}

Finsler metrics can be generated if instead of the Lorentz transforms in SR
there are considered nonlinear generalizations, restrictions of symmetries
and/or deformations of the Minkowski metric, see examples in \cite%
{gibbons,stavr1}. We provide a simple construction when anisotropic metrics
are used for modeling light propagation in anisotropic media (aether) and/or
for small perturbations in quantum gravity.

In SR, a Minkovski metric $\eta _{ij}=diag[-1,+1,+1,+1],$ (for $i=1,2,3,4),$
defines a quadratic line element,$ds^{2}=\eta
_{ij}dx^{i}dx^{j}=-(dx^{1})^{2}+(dx^{2})^{2}+(dx^{3})^{2}+(dx^{4})^{2}.$ The
light velocity $c$ is contained in $x^{1}=ct,$ where $t$ is the time like
coordinate. Along a light ray $x^{i}(\varsigma ),$ parameterized by a real
smooth parameter $0\leq \varsigma \leq \varsigma _{0},$ when $%
ds^{2}/d\varsigma ^{2}=0,$ we can define a ''null'' tangent vector field $%
y^{i}(\varsigma )=dx^{i}/d\varsigma ,$ with $d\tau =dt/d\varsigma .$ Under
general coordinate transforms $x^{i^{\prime }}=x^{i^{\prime }}(x^{i}),$ we
have $\eta _{ij}\rightarrow g_{i^{\prime }j^{\prime }}(x^{k});$ the
condition $ds^{2}/d\varsigma ^{2}=0$ holds always for propagation of light,
i.e. $g_{i^{\prime }j^{\prime }}y^{i^{\prime }}y^{j^{\prime }}=0.$ We can
write for some classes of coordinate systems $c^{2}=g_{\widehat{i}\widehat{j}%
}(x^{i})y^{\widehat{i}}y^{\widehat{j}}/\tau ^{2}$ (for simplicity, omitting
priming of indices and writting $\widehat{i},\widehat{j},...=2,3,4)$.
\footnote{%
This formula holds also in GR, when the local coordinates on a (pseudo)
Riemannian manifold are chosen such a way that the coefficients of metric $%
g_{\widehat{i}\widehat{j}}(x^{i})$ are constrained to be a solution of
Einstein equations. For certain local constructions in vicinity of a point $%
x_{(0)}^{i}$, we can omit the explicit dependence on $x^{i}$ and consider
only formulas derived for $y^{i}.$}

In anisotropic media (and/or modeling spacetime as an aether model), we can
generalize the quadratic expression $g_{\widehat{i}\widehat{j}}(x^{i})y^{%
\widehat{i}}y^{\widehat{j}}$ to an arbitrary nonlinear one $\check{F}^{2}(y^{%
\widehat{j}})$ subjected to the condition of homogeneity that $\check{F}%
(\beta y^{\widehat{j}})=\beta \check{F}(y^{\widehat{j}}),$ for any $\beta
>0. $ The formula for light propagation transforms into $c^{2}=\check{F}%
^{2}(y^{\widehat{j}})/\tau ^{2}.$ Small deformations of the Minkowski metric
can be parameterized in the form $\check{F}^{2}(y^{\widehat{j}})\approx
\left( \eta _{\widehat{i}\widehat{j}}y^{\widehat{i}}y^{\widehat{j}}\right)
^{r}+q_{\widehat{i}_{1}\widehat{i}_{2}...\widehat{i}_{2r}}y^{\widehat{i}%
_{1}}...y^{\widehat{i}_{2r}},$ for $r=1,2,.... $ and $\widehat{i}_{1},%
\widehat{i}_{2},...,\widehat{i}_{2r}=2,3,4. $ In the approximations $r=1$
and $q_{\widehat{i}_{1}\widehat{i}_{2}...\widehat{i}_{2r}}\rightarrow 0,$ we
get the Minkowski (pseudo--Euclidean) spacetime in SR. We can generalize the
coefficients of $\check{F}^{2}$ by introducing additional dependencies on $%
x^{i},$ when $\check{F}^{2}(x^{i},y^{\widehat{j}})\approx \left( g_{\widehat{%
i}\widehat{j}}(x^{k})y^{\widehat{i}}y^{\widehat{j}}\right) ^{r}+q_{\widehat{i%
}_{1}\widehat{i}_{2}...\widehat{i}_{2r}}(x^{k})y^{\widehat{i}_{1}}...y^{%
\widehat{i}_{2r}},$ and consider certain generalized nonlinear homogeneous
relations (with $F(x^{i},\beta y^{j})=\beta F(x^{i},y^{j}),$ for any $\beta
>0$), when {\small
\begin{equation}
ds^{2} = F^{2}(x^{i},y^{j}) \approx -(cdt)^{2}+g_{\widehat{i}\widehat{j}%
}(x^{k})y^{\widehat{i}}y^{\widehat{j}}[1+\frac{1}{r}\frac{q_{\widehat{i}_{1}%
\widehat{i}_{2}...\widehat{i}_{2r}}(x^{k})y^{\widehat{i}_{1}}...y^{\widehat{i%
}_{2r}}}{\left( g_{\widehat{i}\widehat{j}}(x^{k})y^{\widehat{i}}y^{\widehat{j%
}}\right) ^{r}}] +O(q^{2}).  \label{fbm}
\end{equation}
} A nonlinear element $ds^{2}=F^{2}(x^{i},y^{j})$ is usually called by
physicists a ''Finslerian metric''. In the bulk of geometric constructions
in books \cite{ma,bcs}, the function $F$ is considered to be a fundamental
(and/or generating) Finsler function satisfying the condition that the
Hessian $\ ^{F}g_{ij}(x^{i},y^{j})=\frac{1}{2}\frac{\partial F^{2}}{\partial
y^{i}\partial y^{j}}$ is not degenerate.\footnote{%
It is positively definite for models of Finsler geometry; for
pseudo--Finsler configurations \cite{ijgmmp,visser,vgensol,vcosm1}, this
condition is not imposed. Here we also note that physical implications of
Finsler type deformations of SR were analyzed in \cite%
{beken,perlick,mignemi,gibbons,stavr1}.}

\subsubsection{Nonlinear dispersion relations}

The nonlinear quadratic element (\ref{fbm}) results in a nonlinear \
dispersion relation between the frequency $\omega $ and the wave vector $%
k_{i}$ of a light ray (see details, for instance, in \cite{lammer}),%
\begin{equation}
\omega ^{2}=c^{2}\left[ g_{\widehat{i}\widehat{j}}k^{\widehat{i}}k^{\widehat{%
j}}\right] ^{2}\left( 1-\frac{1}{r}\frac{q_{\widehat{i}_{1}\widehat{i}_{2}...%
\widehat{i}_{2r}}y^{\widehat{i}_{1}}...y^{\widehat{i}_{2r}}}{[ g_{\widehat{i}%
\widehat{j}}k^{\widehat{i}}k^{\widehat{j}}] ^{2r}}\right) ,  \label{disp}
\end{equation}%
where, for simplicity, we consider such a relation in a fixed point \ $%
x^{k}=x_{(0)}^{k},$ \ when $g_{\widehat{i}\widehat{j}}(x_{0}^{k})=g_{%
\widehat{i}\widehat{j}}$ and $q_{\widehat{i}_{1}\widehat{i}_{2}...\widehat{i}%
_{2r}}=q_{\widehat{i}_{1}\widehat{i}_{2}...\widehat{i}_{2r}}$ $(x_{0}^{k}).$
The coefficients $q_{\widehat{i}_{1}\widehat{i}_{2}...\widehat{i}_{2r}}$
should be computed from a model of quantum gravity, or from a well defined
Finsler like modification of the general relativity theory. For a locally
anisotropic spacetime aether, i.e. in a modified classical model of gravity
with broken local Lorentz invariance, the coefficients for dispersions of
type (\ref{disp}) have be measured following some experiments when light
rays propagate according to a Riemannian/ Finsler metric.

Dispersion relations should be parameterized and computed differently for
theories with nonlocal interactions and noncommutative variables.
Nevertheless, the form (\ref{disp}) is a very general one which can be
obtained in various Finsler like and extra dimension models even the values
of coefficients $q_{\widehat{i}_{1}\widehat{i}_{2}...\widehat{i}_{2r}}$
depend on the class of exact solutions of certain generalized gravitational
equations, \ types of classical and quantum models etc.

\subsubsection{Finsler metrics do not define complete geometric models}

Any (pseudo) Riemannian geometry on (for our purposes, we consider any
necessary smooth class) manifold $M$ is determined by a metric field $%
g=g_{ij}(x)e^{i}\otimes e^{i},$ where $x=\{x^{i}\}$ label the local
coordinates and the coefficients of a symmetric tensor $g_{ij}(x)$ are
defined with respect to a general nonholonomic co-frame $e^{i}=e_{\
\underline{i}}^{i}(x)dx^{\underline{i}}.$\footnote{%
We follow the system of notations from \cite{vrflg,vsgg,ijgmmp,vncg} when
''underlined'', ''primed'' and other type indices are used in order to
distinguish, for instance, the local coordinate co-base $dx^{\underline{i}}$%
, with $dx^{\underline{i}}dx^{\underline{j}}-dx^{\underline{j}}dx^{%
\underline{i}}=0,$ and an arbitrary one, $e^{i},$ for which certain
nonholonomy (equivalently, anholonomy, or non--integrability) conditions are
satisfied, $e^{i}e^{j}-e^{j}e^{i}=w_{\ k}^{ij}e^{k},$ with $w_{\ k}^{ij}$
being the anholonomy coefficients. For simplicity, we shall omit
priming/underling of indices if that will not result in ambiguities.} There
is on $M$ a second fundamental geometric object, the Levi--Civita
connection, $\nabla =\{\nabla _{i}\}$ (parameterized locally by coefficients
$\ _{\shortmid }\Gamma _{jk}^{i}$ for a 1--form $\ _{\shortmid }{\Gamma }%
_{j}^{i}=\ _{\shortmid }\Gamma _{jk}^{i}(x)dx^{k})$ which is completely
defined by a set $\{g_{ij}\}$ if and only if we impose two basic conditions:
1) metric compatibility, $\nabla _{k}g_{ij}=0$; 2) zero torsion, $\
_{\shortmid }\mathcal{T}^{i}=\nabla e^{i}=de^{i}+\ _{\shortmid }{\Gamma }%
_{j}^{i}\wedge e^{j}=0$, where $\wedge $ is the anti--symmetric product
forms (see, for instance, \cite{misner}).

Contrary to (pseudo) Riemannian geometry completely determined by a
quadratic linear form (a metric), a Finsler metric (\ref{fbm}) does not
state a geometric spacetime model in a self--consistent and complete form.
An element $ds=F(x^{i},y^{j})$ and Hessian $\ ^Fg_{ij}(x^{i},y^{j})$ do not
define completely any metric and connections structures on the total space $%
TM$ of a tangent bundle $\left( TM,\pi ,M\right),$ where $\pi $ is a
surjective projection (see \cite{ma,bcs}).

There are necessary additional suppositions in order to elaborate a
"well--defined" spacetime model generated by $F(x^{i},y^{j}),$ i.e. a
Finsler gravity theory. Such a locally anisotropic gravity is determined by
three fundamental geometric objects on $TM$ and $TTM,$ a metric structure, $%
^{F}\mathbf{g},$ a nonlinear connection, $^{F}\mathbf{N},$ and a linear
connection, $^{F}\mathbf{D},$ which is adapted to $^{F}\mathbf{N}.$\footnote{%
We put a left label $F$ in order to emphasize (if necessary) that some
objects are introduced for a Finsler geometry model.}

\subsection{Finsler geometry on nonholnomic spacetimes}

\label{ssfgo} We consider the main concepts and fundamental geometric
objects which are necessary for Finsler geometry, and gravity, models on
nonholonomic tangent bundles/ manifolds (spacetimes).

\subsubsection{Fundamental geometric objects in Finsler geometry}

\paragraph{Nonlinear connections:}

\quad A rigorous analysis of the nonlinear connection structures $\ ^{F}%
\mathbf{N}$ was many times omitted by physicists in their works on Finsler
gravity and and analyzes of experimental restrictions on ''velocity''
dependent theories \cite{beken,will}. This geometric/ physical object is
less familiar to researches working in particle physics and cosmology and it
is confused with nonlinear realizations of connections for generalized gauge
theories.

A nonlinear connection (N--connection) $\mathbf{N}$ can be defined as a
Whitney sum (equivalently, a nonholonomic distribution)%
\begin{equation}
TTM=hTM\oplus vTM,  \label{ws1}
\end{equation}%
with a conventional splitting into horizontal (h), $hTM,$ and vertical (v), $%
vTM,$ subspaces. It is given locally by a set of coefficients $\mathbf{N=}%
\left\{ N_{i}^{a}\right\} ,$ when\footnote{%
coordinates $u=(x,y)$ on an open region $U\subset T\mathbf{M}$ are labelled
in the form $u^{\alpha }=(x^{i},y^{a}),$ with indices of type $%
i,j,k,...=1,2,...n$ and $a,b,c...=n+1,n+2,...,n+n;$ on $TM,$ $x^{i}$ and $%
y^{a}$ are respectively the base coordinates and fiber (velocity like)
coordinates;\ we use boldface symbols for spaces (and geometric objects on
such spaces) enabled with N--connection structure} $\mathbf{N}%
=N_{i}^{a}(u)dx^{i}\otimes \frac{\partial }{\partial y^{a}}$. There is a
frame (vielbein) structure which is linear on N--connection coefficients and
on partial derivatives $\partial _{i}=\partial /\partial x^{i}$ and $%
\partial _{a}=\partial /\partial y^{a}$ and, respectively, theirs duals, $%
dx^{i}$ and $dy^{a},$
\begin{eqnarray}
\mathbf{e}_{\nu } &=&\left( \mathbf{e}_{i}=\partial _{i}-N_{i}^{a}\partial
_{a},e_{a}=\partial _{a}\right) ,  \label{dder} \\
\mathbf{e}^{\mu } &=&\left( e^{i}=dx^{i},\mathbf{e}%
^{a}=dy^{a}+N_{i}^{a}dx^{i}\right) .  \label{ddif}
\end{eqnarray}%
The vielbeins (\ref{ddif}) satisfy the nonholonomy relations $\lbrack
\mathbf{e}_{\alpha },\mathbf{e}_{\beta }]=\mathbf{e}_{\alpha }\mathbf{e}%
_{\beta }-\mathbf{e}_{\beta }\mathbf{e}_{\alpha }=w_{\alpha \beta }^{\gamma }%
\mathbf{e}_{\gamma }$ with (antisymmetric) nontrivial anholonomy
coefficients $w_{ia}^{b}=\partial _{a}N_{i}^{b}$ and $w_{ji}^{a}=\Omega
_{ij}^{a},$ where $\Omega _{ij}^{a}=\mathbf{e}_{j}\left( N_{i}^{a}\right) -%
\mathbf{e}_{i}\left( N_{j}^{a}\right)$ are the coefficients of N--connection
curvature.\footnote{%
The holonomic/ integrable frames are selected by the integrability
conditions $w_{\alpha \beta }^{\gamma }=0.$}

For a $TM$ endowed with a generating Finsler function $F$, we can introduce
a homogeneous Lagrangian $L=F^{2}.$ There is the canonical (Cartan's)
N--connection with coefficients
\begin{equation}
\ ^{c}N_{i}^{a}=\frac{\partial G^{a}}{\partial y^{n+i}},\ \mbox{ for } \
G^{a}=\frac{1}{4}\ \ ^{F}g^{a\ n+i} \left( \frac{\partial ^{2}L}{\partial
y^{n+i}\partial x^{k}}y^{n+k} -\frac{\partial L}{\partial x^{i}}\right),
\label{clnc}
\end{equation}%
where $\ ^{F}g^{ab}$ is inverse to $\ ^{F}g_{ab}$.\footnote{%
Respective contractions of $h$-- and $\ v$--indices, $\ i,j,...$ and $%
a,b..., $ are performed following the rule: we write an up $v$--index $a$ as
$a=n+i$ and contract it with a low index $i=1,2,...n;$ on total spaces of
even dimensions, we can write $y^{i}$ instead of $y^{n+i},$ or $y^{a}.$ The
spacetime signature may be encoded formally into certain systems of frame
(vielbein) coefficients and coordinates, some of them being proportional to
the imaginary unity $i,$ when $i^{2}=-1.$ For a local tangent Minkowski
space of signature $(-,+,+,+),$ we can chose $e_{0^{\prime }}=i\partial
/\partial u^{0^{\prime }},$ where $i$ is the imaginary unity, $i^{2}=-1,$
and write $e_{\alpha ^{\prime }}=(i\partial /\partial u^{0^{\prime
}},\partial /\partial u^{1^{\prime }},\partial /\partial u^{2^{\prime
}},\partial /\partial u^{3^{\prime }}).$ Euclidean coordinates with $i$ were
used in textbooks on relativity theory (see, for instance, \cite%
{landau,moller}). Latter, they were considered for analogous modelling of
gravity theories as effective Lagrange mechanics, or Finsler like,
geometries \cite{vrflg,ijgmmp}. The term ''pseudo--Finsler'' was also
introduced in a different form, more recently, for some analogous gravity
models (see, for instance, \cite{visser}) and a mathematical book \cite%
{bejf1}.}

For any set $N_{i}^{a},$ we can chose a well--defined $F$ and corresponding
frame coefficients $\ e_{\ \alpha ^{\prime }}^{\alpha }$ and (inverse) $%
e_{\alpha }^{\ \alpha ^{\prime }}$ when \ $e_{\alpha }\rightarrow e_{\alpha
}^{\ \alpha ^{\prime }}$ $e_{\alpha ^{\prime }}$ transform $N_{i}^{a}$ into
respective $\ ^{c}N_{i}^{a}$ (for instance, $N_{i}^{a}=e_{i}^{\ i^{\prime
}}e_{\ a^{\prime }}^{a}\ ^{c}N_{i}^{a},$ or using more general types of
transforms). So, we can work equivalently with any convenient set $\mathbf{N}%
=\{N_{i}^{a}\}$ which may be redefined as a ''canonical'' set $\
^{F}N_{i}^{a}=\ ^{c}N_{i}^{a}.$

In our works devoted to applications of Finsler geometry methods in modern
gravity and string theory \cite{vrflg,vsgg,ijgmmp} related to standard
physics models, we considered N--connections not only on tangent bundles but
also on nonholonomic manifolds (see definition in footnote \ref{fnnhm}). In
such approaches, it is considered a general manifold $\mathbf{V,}$ instead
of $TM,$ when the Whitney sum (\ref{ws1}), existing naturally on
vector/tangent bundles, is introduced as a nonholonomic distribution on $%
\mathbf{V}$ with conventional h-- and v--splitting into (holonomic and
nonholonomic variables, respectively, distinguished by coordinates $x^{i}$
and $y^{a}),$%
\begin{equation}
T\mathbf{V}=h\mathbf{V}\oplus v\mathbf{V}.  \label{whitney}
\end{equation}

A N--anholonomic manifold (or tangent bundle; in brief, we shall write
respectively the terms "bundle" and "manifold"; we can consider similarly
vector bundles) is a nonholonomic manifold enabled with N--connection
structure (\ref{whitney}). The properties of a N--anholono\-mic
bundle/manifold are determined by N--adapted bases (\ref{dder}) and (\ref%
{ddif}). A geometric object is N--adapted (equivalently, distinguished),
i.e. it is a d--object, if it can be defined by components adapted to the
splitting (\ref{whitney}) (one uses terms d--vector, d--form, d--tensor).
For instance, a d--vector is represented as $\mathbf{X}=X^{\alpha }\mathbf{e}%
_{\alpha }=X^{i}\mathbf{e}_{i}+X^{a}e_{a}$ and a one d--form $\widetilde{%
\mathbf{X}}$ (dual to $\mathbf{X}$) is represented as $\widetilde{\mathbf{X}}%
=X_{\alpha }\mathbf{e}^{\alpha }=X_{i}e^{i}+X_{a}e^{a}.$\footnote{%
A geometric object can be redefined equivalently for arbitrary frame and
coordinate systems; nevertheless, the N--adapted constructions allow us to
preserve a prescribed h-- and v--splitting. In this paper, we omit details
on coordinate transforms of geometric objects (for instance, for the
coefficients of linear and linear connections), on vector/tangent bundles or
nonholonomic manifolds, which are considered in Refs. \cite%
{ma,vrflg,vsgg,ijgmmp}. Geometrically, all formulas are similar and do not
depend on the fact what type of Whitney sum, we use (\ref{whitney}) or (\ref%
{ws1}). There are differences depending on the type of physical theory we
model. For instance, on $TM,$ the v--components can be related to some
"velocity" components, but on $\mathbf{V}$ such v--components are
distinguished by non--integrable constraints encoded into the nonholonomic
frame structure. The h--v--splitting exists naturally (it can be a holonomic
or anholonomic one, depending on the type of geometric/physical model we
consider) on any vector bundle. Such a splitting can be modelled as a local
fibred structure on any (for instance, pseudo--Riemannian) manifold by
introducing corresponding classes of nonholonomic frames. We can introduce
the h- and v--decompositions in the same way on $T\mathbf{V}$ and/or $TTM$
fixing a corresponding N--connection structure, with very similar rules of
transforms of geometric/physical objects. Nevertheless, physical meaning of
such objects are completely different for constructions on tangent bundles
and nonholonomic manifolds.}

\paragraph{Lifts of base metrics on total spaces:}

\ The most known procedure to extend $\ ^{F}g_{ab}$ to a metric in $TM$ is
the so--called Sasaki type lift when the Hessian metric is considered in
N--adapted form both of the h-- and v--components metric,
\begin{eqnarray}
\ ^{F}\mathbf{g} &=&\ \ ^{F}\mathbf{g}_{\alpha \beta } du^{\alpha }\otimes
du^{\beta } = \ ^{F}g_{ij}\left( u\right) dx^{i}\otimes dx^{j}+\
^{F}g_{ab}\left( u\right) \ ^{c}\mathbf{e}^{a}\otimes \ ^{c}\mathbf{e}^{b},
\notag \\
\ ^{c}\mathbf{e}^{a} &=&dy^{a}+\ ^{c}N_{i}^{a}\left( u\right) dx^{i},\
du^{\alpha }=\left( dx^{i},dy^{a}\right).  \label{fsm}
\end{eqnarray}%
Similarly, we can define a metric structure for an even dimensional
N--anholonomic manifold $\mathbf{V}$ (this condition is satisfied for any $%
TM)$\ endowed with a h--metric $g_{ij},$ on $h\mathbf{V,}$ and a given set
of N--connection coefficients $N_{i}^{a}.$

Using arbitrary frame transforms with coefficients $e_{\ \alpha ^{\prime
}}^{\alpha }(u),$ we can transform the total Finsler metric (\ref{fsm}) into
a ''general'' one $\mathbf{g=g}_{\alpha ^{\prime }\beta ^{\prime }}\mathbf{e}%
^{\alpha ^{\prime }}\otimes \mathbf{e}^{\beta ^{\prime }}$ on $TM,$ where $%
\mathbf{g}_{\alpha ^{\prime }\beta ^{\prime }}=e_{\ \alpha ^{\prime
}}^{\alpha }e_{\ \beta ^{\prime }}^{\beta }\ \ ^{F}\mathbf{g}_{\alpha \beta
} $ and $\mathbf{e}^{\alpha ^{\prime }}=e_{\alpha }^{\ \alpha ^{\prime
}}\left( u\right) \ ^{c}\mathbf{e}^{\alpha },$ for $\ ^{c}\mathbf{e}^{\alpha
}=(dx^{i},\ ^{c}\mathbf{e}^{a}).$ A metric d--tensor (d--metric) is with $%
n+n $ splitting. Such a N--adapted decomposition can be performed by
corresponding parametrizations of components of matrices $e_{\ \alpha
^{\prime }}^{\alpha },$ when $\mathbf{g}_{\alpha ^{\prime }\beta ^{\prime
}}=[g_{ij},h_{ab},N_{i}^{a}].$ Haven redefined the coordinates and frame
coefficients, we can express
\begin{eqnarray}
\ \ \mathbf{g} &=&\ g_{ij}(x,y)dx^{i}\otimes dx^{j}+h_{ab}(x,y)\mathbf{e}%
^{a}\otimes \mathbf{e}^{b},  \label{gdm} \\
\ \mathbf{e}^{a} &=&dy^{a}+N_{i}^{a}(x,y)dx^{i}.  \notag
\end{eqnarray}
With respect to local dual coordinate frames, a metric (\ref{gdm}) is
parameterized
\begin{eqnarray}
\mathbf{g}&=&\ \underline{g}_{\alpha \beta }\left( u\right) du^{\alpha
}\otimes du^{\beta },  \label{fmetr} \\
\underline{g}_{\alpha \beta } &=&\left[
\begin{array}{cc}
\ g_{ij}+N_{i}^{a}N_{j}^{b}h_{ab} & N_{j}^{e}h_{ae} \\
N_{i}^{e}\ h_{be} & \ h_{ab}%
\end{array}%
\right] .  \label{fansatz}
\end{eqnarray}%
We emphasize that the values $N_{i}^{a}(u)$ should not be identified as
certain gauge fields in a Kaluza--Klein theory if we do not consider
compactifications on coordinates $y^{a}.$\footnote{%
In Finsler like theories, a set $\{N_{i}^{a}\}$ defines a N--connection
structure, with elongated partial derivatives \ (\ref{dder}). In
Kaluza--Klein gravity they are linearized on $y^{a},$ $%
N_{i}^{a}=A_{bi}^{a}(x)y^{b},$ when $A_{bi}^{a}(x)$ are treated as some
Yang--Mills potentials inducing covariant derivatives.}

For simplicity, hereafter we shall work on a general nonholonomic space $%
\mathbf{V}$ enabled with N--connection splitting $\mathbf{N}$ (\ref{whitney}%
) and resulting N--adapted base and co--base (\ref{dder}) and (\ref{ddif}).
Such a space is also endowed with a symmetric metric structure $\mathbf{g}$
(of necessary local Euclidean or pseudo--Euclidean signature) which can be
parameterized in the form (\ref{gdm}) (or, equivalently, (\ref{fmetr})). Any
metric $\mathbf{g}$ on $\mathbf{V}$ can be represented equivalently in the \
form $\ \ ^{F}\mathbf{g}$ (\ref{fsm}) after corresponding frame and
coordinate transforms for a well defined generating function $F(x,y).$ It is
always possible to introduce on a (pseudo) Riemannian manifold/ tangent
bundle $\mathbf{V}$ some local Finsler like variables when the metric is
parametrezid in a Sasaki type form. We shall write $\ \mathbf{V=}TM\ \ $when
it will be necessary to emphasize that the constructions are defined
explicitly for tangent bundles.

\paragraph{Distinguished connections, theirs torsions and curvatures:}

\quad For any d--metric of type (\ref{fsm}) and/or (\ref{gdm}), we can
construct the Levi--Civita connection $\nabla =\{\ _{\shortmid }\Gamma _{\
\beta \gamma }^{\alpha }\}$ on $\mathbf{V}$ in a standard form.
Nevertheless, this connection is not used in Finsler geometry and
generalizations. The problem is that $\nabla $ is not compatible with a
N--connection splitting, i.e. under parallel transports with $\nabla ,$ it
is not preserved the Whitney sum (\ref{whitney}).

In order to perform geometric constructions with h--/v--splitting, it was
introduced the concept of distinguished connection (in brief,
d--connection). By definition, such a d--connection $\mathbf{D=(}h\mathbf{D,}%
v\mathbf{D)}$ is a linear one \ preserving under parallelism the
N--connection structure on $\mathbf{V.}$ The N--adapted components $\mathbf{%
\Gamma }_{\ \beta \gamma }^{\alpha }$ of a d--connection $\mathbf{D}$ are
defined by equations $\mathbf{D}_{\alpha }\mathbf{e}_{\beta }=\mathbf{\Gamma
}_{\ \alpha \beta }^{\gamma }\mathbf{e}_{\gamma }$ and parameterized in the
form $\ \mathbf{\Gamma }_{\ \alpha \beta }^{\gamma }=\left(
L_{jk}^{i},L_{bk}^{a},C_{jc}^{i},C_{bc}^{a}\right) ,$ where $\mathbf{D}%
_{\alpha }=(D_{i},D_{a}),$ with $h\mathbf{D}=(L_{jk}^{i},L_{bk}^{a})$ and $v%
\mathbf{D}=(C_{jc}^{i},$ $C_{bc}^{a})$ defining the covariant, respectively,
h-- and v--derivatives.

\ The simplest way to perform computations with a d--connection $\mathbf{D}$
is to associate it with a N--adapted differential 1--form $\mathbf{\Gamma }%
_{\ \beta }^{\alpha }=\mathbf{\Gamma }_{\ \beta \gamma }^{\alpha }\mathbf{e}%
^{\gamma },$ when the coefficients of forms and tensors (i.e. d--tensors
etc) are defined with respect to (\ref{ddif}) and (\ref{dder}). In this
case, we can apply the well known formalism of differential forms as in
general relativity \cite{misner}. It also allows us to elaborate an
N--adapted differential/integral calculus for Finsler spaces and
generalizations. For instance, torsion of $\ \mathbf{D}$ is defined/computed
\begin{equation}
\mathcal{T}^{\alpha }\doteqdot \mathbf{De}^{\alpha }=d\mathbf{e}^{\alpha }+%
\mathbf{\Gamma }_{\ \beta }^{\alpha }\wedge \mathbf{e}^{\beta },
\label{tors}
\end{equation}%
see formulas (\ref{dtors}) in Appendix, for explicit values of coefficients $%
\mathcal{T}^{\alpha }=\{\mathbf{T}_{\ \beta \gamma }^{\alpha }\}.$
Similarly, using the d--connection 1--form (\ref{dder}), one computes the
curvature of $\mathbf{D}$ (d--curvature)
\begin{equation}
\mathcal{R}_{~\beta }^{\alpha }\doteqdot \mathbf{D\Gamma }_{\ \beta
}^{\alpha }=d\mathbf{\Gamma }_{\ \beta }^{\alpha }-\mathbf{\Gamma }_{\ \beta
}^{\gamma }\wedge \mathbf{\Gamma }_{\ \gamma }^{\alpha }=\mathbf{R}_{\ \beta
\gamma \delta }^{\alpha }\mathbf{e}^{\gamma }\wedge \mathbf{e}^{\delta },
\label{curv}
\end{equation}%
see formulas (\ref{dcurv}) for h--v--adapted components, $\mathcal{R}%
_{~\beta }^{\alpha }=\{\mathbf{\mathbf{R}}_{\ \ \beta \gamma \delta
}^{\alpha }\}.$

The Ricci d--tensor $Ric=\{\mathbf{R}_{\alpha \beta }\}$ is defined in a
standard form by contracting respectively the components of (\ref{dcurv}), $%
\mathbf{R}_{\alpha \beta }\doteqdot \mathbf{R}_{\ \alpha \beta \tau }^{\tau
},$ The h--/ v--components of this d--tensor, $\mathbf{R}_{\alpha \beta
}=\{R_{ij},R_{ia},\ R_{ai},\ R_{ab}\},$ are
\begin{equation}
R_{ij}\doteqdot R_{\ ijk}^{k},\ \ R_{ia}\doteqdot -R_{\ ika}^{k},\
R_{ai}\doteqdot R_{\ aib}^{b},\ R_{ab}\doteqdot R_{\ abc}^{c},
\label{dricci}
\end{equation}%
see explicit coefficients formulas (\ref{dcurv}). Here we emphasize that for
an arbitrary d--connecti\-on $\mathbf{D,}$ this tensor is not symmetric,
i.e. $\mathbf{R}_{\alpha \beta }\neq \mathbf{R}_{\beta \alpha }.$ In order
to define the scalar curvature of a d--connection $\mathbf{D,}$ we have to
use a d--metric structure $\mathbf{g}$ (\ref{gdm}) on $\mathbf{V,}$ or $TM,$
$\ ^{s}\mathbf{R}\doteqdot \mathbf{g}^{\alpha \beta }\mathbf{R}_{\alpha
\beta }=g^{ij}R_{ij}+h^{ab}R_{ab},$ with $R=g^{ij}R_{ij}$ and $%
S=h^{ab}R_{ab} $ being respectively the h-- and v--components of scalar
curvature.

For any d--connection $\mathbf{D}$ in Finsler geometry, and generalizations,
the Einstein d--tensor is (by definition)
\begin{equation}
\mathbf{E}_{\alpha \beta }\doteqdot \mathbf{R}_{\alpha \beta }-\frac{1}{2}%
\mathbf{g}_{\alpha \beta }\ ^{s}\mathbf{R}.  \label{enstdt}
\end{equation}%
This d--tensor is also not symmetric and, in general, $\mathbf{D}_{\alpha }%
\mathbf{E}^{\alpha \beta }\neq 0.$ Such a tensor is very different from that
for the Levi--Civita connection $\nabla $ which is symmetric and with zero
covariant divergence, i.e. $E_{\alpha \beta }=E_{\beta \alpha }$ and $\nabla
_{\alpha }E^{\alpha \beta }\neq 0,$ where $E_{\alpha \beta }$ is computed
using $\ _{\shortmid }\Gamma _{\ \beta \gamma }^{\alpha }.$

\subsubsection{Notable connections for Finsler spaces}

In general, it is possible to define on $\mathbf{V}$ two independent
fundamental geometric structures $\mathbf{g}$ and $\mathbf{D}$ which are
adapted to a given $\mathbf{N}$. For applications in modern physics, it is
more convenient to work with a d--connection $\mathbf{D}$ which is metric
compatible satisfying the condition $\mathbf{Dg}=0,$ see discussions in \cite%
{vrflg,vsgg,ijgmmp}. There is an infinite number of d--connections which are
compatible to a metric $\mathbf{g}.$ A special interest presents a subclass
of such metrics which are completely defined by $\mathbf{g}$ in a unique
N--adapted form following a well defined geometric principle.

\paragraph{The canonical d--connection:}

\quad In our works on Finsler gravity, we used the so--called canonical
d--connection $\widehat{\mathbf{D}}$ (on spaces of even dimensions it is
called the $h$--/$v$--connection, such connections were studied
geometrically in \cite{ma} on vector/tangent bundles and for generalized
Finsler geometry).

By definition, $\widehat{\mathbf{D}}$ is with vanishing horizontal and
vertical components of torsion and satisfies the conditions $\widehat{%
\mathbf{D}}\mathbf{g}=0,$ see explicit component formulas (\ref{candcon})
and (\ref{dtors}). \ From many points of view, on a nonholonomic space $%
\mathbf{V,}$ $\widehat{\mathbf{D}}$ is the ''best'' N--adapted analog of the
Levi--Civita connection $\nabla .$ We have the distortion relation
\begin{equation}
\nabla =\widehat{\mathbf{D}}+\widehat{\mathbf{Z}},  \label{distorsrel}
\end{equation}%
when both linear connections $\nabla =\{\ _{\shortmid }\Gamma _{\ \beta
\gamma }^{\alpha }\}$ and $\widehat{\mathbf{D}}=\{\widehat{\mathbf{\Gamma }}%
_{\ \alpha \beta }^{\gamma }\}$ and the distorting tensor $\widehat{\mathbf{Z%
}}=\{\widehat{\mathbf{\ Z}}_{\ \alpha \beta }^{\gamma }\}$ are uniquely
defined by the same metric tensor $\mathbf{g.}$ The coefficient formulas are
given in Appendix, see (\ref{deflc}) and (\ref{deft}). The connection $%
\widehat{\mathbf{D}}$ is with nontrivial torsion (the coefficients $\widehat{%
T}_{\ ja}^{i},\widehat{T}_{\ ji}^{a}$ and $\widehat{T}_{\ bi}^{a}$ are not,
in general, zero, see (\ref{dtors})) but such d--torsions are
nonholnomically induced by N--connection coefficients and completely
determined by certain off--diagonal N--terms in (\ref{fansatz}). All
geometric constructions can be performed equivalently and redefined in terms
of both connections $\nabla $ and $\widehat{\mathbf{D}}$ using (\ref%
{distorsrel}).

The connection $\widehat{\mathbf{D}}$ and various types of d--connections $%
\mathbf{D=(}h\mathbf{D,}v\mathbf{D)}$ with h- and v--covariant derivatives, $%
h\mathbf{D}=(L_{jk}^{i},L_{bk}^{a})$ and $v\mathbf{D}=(C_{jc}^{i},$ $%
C_{bc}^{a})$ can be defined on vector bundles and on (pseudo) Riemann spaces
of arbitrary dimensions, alternatively to $\nabla$. A general Finsler
d--connection is of type $\ ^{F}\mathbf{D=\{}L_{jk}^{i},C_{jc}^{i}\mathbf{\}}
$ with coefficients $L_{jk}^{i}$ and $C_{jc}^{i}$ determined by a generating
Finsler function $F$ and a N--connection $\mathbf{N=}\left\{
N_{i}^{a}\right\} $ and some arbitrary and/or induced nonholonmically
torsion fields. In general, $\ ^{F}\mathbf{D}$ is not metric compatible,
i.e. $\ ^{F}\mathbf{D}\ ^{F}\mathbf{g=\ ^{F}Q}\neq 0.$

\paragraph{Nonmetric Finsler d--connections:}

\ There were considered three types of such notable connections (see details
and references in \cite{ma,bcs} and, on physical applications, \cite%
{vrflg,vsgg}). In general, it is possible to define following different
geometric principles an infinite number of metric noncompatible or
compatible d--connections in Finsler geometry and generalizations. The first
(metric noncompatible) one was the Berwald d--connection $\ ^{F}\mathbf{D=}\
^{B}\mathbf{\mathbf{D}=\{}L_{bk}^{a}=\partial N_{k}^{a}/\partial
y^{b},C_{jc}^{i}=0\mathbf{\}.}$ It is completely defined by the
N--connection structure and $\ ^{B}\mathbf{D}\ ^{F}\mathbf{g=}\ ^{B}\mathbf{Q%
}\neq 0.$

Then, it was introduced the Chern d--connection \cite{bcs}, $\ ^{F}\mathbf{D}%
=\ ^{Ch}\mathbf{\mathbf{D}}=\ \widehat{L}_{jk}^{i},C_{jc}^{i}=0, $ where $%
\widehat{L}_{jk}^{i}$ is given by the first formula in (\ref{candcon}). This
d--connection is torsionless, $\ ^{Ch}\mathcal{T}=0,$ but (in general)
metric noncompatible, $\ ^{Ch}\mathbf{D}\ ^{F}\mathbf{g}=\ ^{Ch}\mathbf{Q}%
\neq 0.$ Recently, some authors attempted to elaborate Chern--Finsler, or
Berwald--Finsler cosmological models and certain modifications of Einstein
gravity using such metric noncompatible d--connections \cite%
{lichang5,lichang6}. Geometrically, the Chern d--connection is a ''nice''
one but with a number of problems for applications in standard models of
physics (because of nonmetricity, there are difficulties in definition of
spinors and Dirac operators, conservation laws, quantization etc, see
critical remarks in \cite{vcrit,vrflg}). \footnote{%
Experts on Finsler geometry also know about the Hashiguchi d--connection $\
^{F}\mathbf{D=}\ ^{H}\mathbf{\mathbf{D}=\{}L_{bk}^{a}=\partial
N_{k}^{a}/\partial y^{b},\ ^{H}C_{jc}^{i}\mathbf{\},}$ where $\
^{H}C_{jc}^{i}=\frac{1}{2}\ ^{F}g^{ad}( e_{c}\ ^{F}g_{bd}+e_{c}\
^{F}g_{cd}-e_{d}\ ^{F}g_{bc}),$ for $e_{c}=\partial /\partial y^{c}$ and a
given $\ ^{F}g_{bd}$, see details, for instance, in \cite{ma}. It contains
both nontrivial torsion and nonmetricity components, all completely defined
by the N--connection and Hessian structure. We studied Finsler--affine
theories with very general torsion and nonmetric structures, provided
examples of exact solutions and discussed physical implications of models in
Part I of monograph \cite{vsgg}.}

\paragraph{A preferred metric compatible Cartan d--connection:}

\ Historically, it was the first d--connection introduced in Finsler gravity
\cite{cartan} in 1935. Perhaps, the first model of Finsler gravity with the
Einstein equations formulated for the Cartan d--connection was proposed by
J. Horvath \cite{horvath} in 1950. Latter, such constructions were
generalized for other classes of d--connections on vector/tangent bundles
and nonholonomic manifolds, see details in Refs. \cite{ma,vrflg,vsgg}). It
is given by local coefficients $\ ^{F}\mathbf{D}=\ ^{c}\mathbf{\mathbf{D}=\{}%
\ ^{c}L_{ik}^{i},\ ^{c}C_{bc}^{a}\mathbf{\},}$ where {\small
\begin{eqnarray}
\ ^{c}L_{jk}^{i} &=& \frac{1}{2}\ \ ^{F}g^{ir}\left( \mathbf{e}_{k}\
^{F}g_{jr}+\mathbf{e}_{j}\ ^{F}g_{kr}-\mathbf{e}_{r}\ ^{F}g_{jk}\right),
\notag \\
\ \ ^{c}C_{bc}^{a} &=& \frac{1}{2}\ ^{F}g^{ad}\left( e_{c}\
^{F}g_{bd}+e_{c}\ ^{F}g_{cd}-e_{d}\ ^{F}g_{bc}\right).  \label{cartand}
\end{eqnarray}%
} The Cartan d--connection (\ref{cartand}) is metric compatible, $\ ^{c}%
\mathbf{\mathbf{D}}\ ^{F}\mathbf{g}=0,$ but with nontrivial torsion $\ ^{c}%
\mathcal{T}\neq 0$ (the second property follows from formulas (\ref{tors})
and (\ref{dtors}) redefined for $\ ^{c}\mathbf{\mathbf{D).}}$ The nontrivial
torsion terms are induced nonholonomically by a fundamental Finsler function
$F$ via $\ ^{F}g_{bd}$ and Cartan's N--connection $\ ^{c}N_{i}^{a}$ (\ref%
{clnc}). This torsion is very different from the well known torsion in
Einstein--Cartan, or string/gauge gravity models, because in the
Cartan--Finsler case we do not need additional field equations for the
torsion fields. The torsion $\ ^{c}\mathcal{T},$ similarly to $\ \widehat{%
\mathcal{T}},$ is completely defined by a (Finsler) d--metric structure.%
\footnote{%
A very important property of $\ ^{c}\mathbf{\mathbf{D}}$ is that it defines
also a canonical almost symplectic connection \cite{matsumoto,ma}, see
details and recent applications to quantization of Finsler spaces and
Einstein/brane gravity in \cite{vqgrbr}. Perhaps, the Cartan d--connection
is the ''best'' one for physical applications in modern physics of Finsler
geometry and related anholonomic deformation method, see additional
arguments in \cite{vcrit,vrflg,vsgg,vqgrbr}. The d--connections $\ ^{c}%
\mathbf{\mathbf{D}}$ and $\ \widehat{\mathbf{\mathbf{D}}}$ allow us to work
in N--adapted form in Finsler classical and quantum gravity theories keeping
all geometric and physical constructions to be very similar to those for the
Levi--Civita connection $\nabla .$}

In Finslerian theories of gravity, it is possible to work geometrically with
very different types of d--connections because there are always some
transformations of the Cartan d--connection into any mentioned above (or
more general ones) notable Finsler connections, of Berwald, Chern or
Hashiguchi types. Nevertheless, the main issue is that for what kind of
d--connections we can formulate Einstein/Dirac/ Yang--Mills etc equations
which are well defined, self--consistent and with important physical
implications. In our approach, for gravity models on nonholonomic
manifolds/bundles of arbitrary dimension, we give priority to the canonical
d--connection $\widehat{\mathbf{\mathbf{D}}}$ (here we note also that the
Einstein equations for this d--connection can be integrated in very general
forms, \cite{vgensol,vncg,ijgmmp,vsgg,vcosm1}).

\section{Field Equations for Finsler Theories of Gravity}

\label{sfeftg} In this section, we outline the theory of Einstein--Finsler
spaces. There are formulated the Einstein equations for the canonical and/
or Cartan d--connections. It is analyzed a class of theories extending the
four dimensional general relativity to metric compatible Finsler gravities
on tangent bundles and/or nonholonomic manifolds. We end with a discussion
of principles of GR and their extension to metric compatible Finsler gravity
theories.

\subsection{Einstein equations for distinguished connections}

\subsubsection{Gravitational field equations in h--/v--components:}

Having prescribed a N--connection $\mathbf{N}$ and d--metric $\mathbf{g}$ (%
\ref{fsm}) structures on a N--anholonomic manifold $\mathbf{V,}$ for any
metric compatible d--connection $\mathbf{D},$ we can compute the Ricci $%
\mathbf{R}_{\alpha \beta }$ (\ref{dricci}) and Einstein $\mathbf{E}_{\alpha
\beta }$ (\ref{enstdt}) d--tensors. The N--adapted gravitational field
equations \cite{ma,vncg,vsgg,vrflg} are $\mathbf{E}_{\alpha \beta }=\mathbf{%
\Upsilon }_{\alpha \beta },$ where the source $\mathbf{\Upsilon }_{\alpha
\beta }$ has to be defined in explicit form following certain explicit
models of ''locally anisotropic'' gravitational and matter field
interactions.\footnote{%
For theories with arbitrary torsions $\mathcal{T},$\ we have to complete
such equations with additional algebraic or dynamical ones (for torsion's
coefficients) like in the Einstein--Cartan, gauge, or string theories with
torsion. On generalized Finsler spaces, such constructions should be in
N--adapted forms (see details in Part I of \cite{vsgg}). In brief, we note
here that the N--adapted tensor, covariant differential/integral calculus
can be performed very similarly to the well known tetradic formalism, in our
case, using respectively, the N--elongated partial derivatives and
differentials, (\ref{dder}) and (\ref{ddif}). This way, we can elaborate a
N--adapted variational calculus on $TM$, or $\mathbf{V}$ using the
corresponding d--connection and d--metric structures. For metric compatible
d--connections, in N--adapted bases, all constructions are very similar to
those in GR. We can provide proofs for locally anisotropic fluid/spinning
models, Dirac equations, Yang--Mills fields on Finsler spaces etc, see
details in Refs. \cite{vrflg,vsgg}. This is the priority of the
canonical/Cartan d--connection structure. We can not generate "simple"
physical theories if we work with the metric noncompatible Chern
d--connection, see additional critics and discussion in Refs. \cite%
{vcrit,vncbh}.}

The field equations for metric compatible d--connections in Finsler gravity
theories can be distinguished in the form
\begin{eqnarray}
R_{ij}-\frac{1}{2}(R+S)g_{ij} &=&\mathbf{\Upsilon }_{ij},  \label{eq1} \\
R_{ab}-\frac{1}{2}(R+S)h_{ab} &=&\mathbf{\Upsilon }_{ab},  \label{eq2} \\
R_{ai} =\mathbf{\Upsilon }_{ai},\ R_{ia} &=& -\mathbf{\Upsilon }_{ia},
\label{eq4}
\end{eqnarray}%
where $R_{ai}=R_{\ aib}^{b}$ and $R_{ia}=R_{\ ikb}^{k}$ are defined by
formulas for d--curvatures (\ref{dcurv}) containing \ d--torsions (\ref%
{dtors}). For a metric compatible d--connection $\mathbf{D}$ which is
completely defined by a d--metric structure $\mathbf{g,}$ the corresponding
system (\ref{eq1})--(\ref{eq4}) is very similar to that for the usual
Einstein gravity. The difference is that $R_{ai}\neq R_{ia},\nabla \neq
\mathbf{D}$.\footnote{%
Elaborating geometric/gravity models on $TM,$ containing in the limit $%
\mathbf{D\rightarrow }\nabla $ the Einstein gravity theory on $M,$ we should
consider that equations (\ref{eq1}) define a generalization of $\
_{\shortmid }R_{ij}-\frac{1}{2}\ _{\shortmid }Rg_{ij}=\ _{\shortmid
}\Upsilon _{ij}$ for $\nabla =\{\ _{\shortmid }\Gamma _{\ jk}^{i}\}$ and a
well defined procedure of ''compactification'', or brane like
warping/trapping on $y^{a}.$ The observed three dimensional space, with
possible Finsler type contributions, is contained in such classes of
solutions.}

\subsubsection{Equations equations for $\protect\widehat{\mathbf{\mathbf{D}}}
$ and $\ ^{c}\mathbf{\mathbf{D}}$}

Because the canonical d--connection $\widehat{\mathbf{\mathbf{D}}}$ is
completely defined by $\mathbf{g}_{\beta \delta },$ the corresponding
Finsler analog of Einstein, we use the tensor (\ref{enstdt}) for $\mathbf{D}=%
\widehat{\mathbf{\mathbf{D}}},$
\begin{equation}
\widehat{\mathbf{R}}_{\ \beta \delta }-\frac{1}{2}\mathbf{g}_{\beta \delta
}\ ^{s}R=\widehat{\mathbf{\Upsilon }}_{\beta \delta },  \label{ensteqcdc}
\end{equation}%
can be constructed to be equivalent to the Einstein equations for $\nabla .$
This is possible if\ $\ \widehat{\mathbf{\Upsilon }}_{\beta \delta }=\
^{matter}\mathbf{\Upsilon }_{\beta \delta }+\ ^{z}\mathbf{\Upsilon }_{\beta
\delta }$ are derived in such a way that they contain contributions from\ 1)
\ the N--adapted energy--momentum tensor (defined variationally following
the same principles as in general relativity but on $\mathbf{V}$) and 2) the
distortion of the Einstein tensor in terms of $\ \widehat{\mathbf{Z}}$ (\ref%
{distorsrel}), i.e. (\ref{deflc}), $\widehat{\mathbf{E}}_{\ \beta \delta }=\
_{\shortmid }E_{\alpha \beta }+\ ^{z}\widehat{\mathbf{E}}_{\ \beta \delta },$
for $\ ^{z}\widehat{\mathbf{E}}_{\ \beta \delta }=\ ^{z}\mathbf{\Upsilon }%
_{\beta \delta }.$ The value $\ ^{z}\widehat{\mathbf{E}}_{\ \beta \delta }$
is computed by introducing $\widehat{\mathbf{D}}=\nabla -\widehat{\mathbf{Z}}
$ into (\ref{dcurv}).

The system of equations (\ref{dcurv}) can be integrated in very general
forms, see explicit constructions in subsection \ref{sseinst}. Such
solutions can be considered also in general relativity if we impose
additionally the condition that
\begin{equation}
\widehat{L}_{aj}^{c}=e_{a}(N_{j}^{c}),\ \widehat{C}_{jb}^{i}=0,\ \Omega _{\
ji}^{a}=0,  \label{lccond}
\end{equation}%
for $\mathbf{\Upsilon }_{\beta \delta }\rightarrow \varkappa T_{\beta \delta
}$ (matter energy--momentum in Einstein gravity)\ if $\widehat{\mathbf{D}}%
\rightarrow \nabla .$ \ We emphasize here that if the constraints (\ref%
{lccond}) are satisfied the tensors $\widehat{\mathbf{T}}_{\ \alpha \beta
}^{\gamma }$(\ref{dtors}) and $Z_{\ \alpha \beta }^{\gamma }$(\ref{deft})
are zero. For such configurations, we have $\widehat{\mathbf{\Gamma }}_{\
\alpha \beta }^{\gamma }=\ _{\shortmid }\Gamma _{\ \alpha \beta }^{\gamma },$
with respect to (\ref{dder}) and (\ref{ddif}), see (\ref{deflc}), even $%
\widehat{\mathbf{D}}\neq \nabla .$

There are two very important benefits to work with canonical Finsler
variables (for instance with the Cartan d--connection) on (pseudo)
Riemannian manifolds: 1) the Einstein equations "magically" separate and can
be integrated in very general forms \cite{vgensol,ijgmmp,vncg}; 2) there is
also an equivalent almost K\"{a}hler representation for such Finsler
variables in GR which allows us to perform various types of
deformation/A--brane quantization and/or two connection renormalization \cite%
{vegpla,vqgrbr,vqggr,vcrit}. In Finsler geometry/gravity models, the
constraints (\ref{lccond}) are not obligatory. On $TM,$ and any even
dimensional $\mathbf{V,}$ it is possible to perform such frame deformations
when $\widehat{\mathbf{D}}\rightarrow $ $\ ^{c}\mathbf{\mathbf{D}}$. So, the
Einstein equations for the Cartan d--connection, in GR and Finsler
generalizations, also can be integrated in very general forms.

An extra dimensional gravity theory can be elaborated for a linear
connection (in general, it can be metric noncompatible) with $y^a$ are
considered as "extra dimension" coordinates to a 4--d (pseudo) Riemannian
spacetime manifold with coordinates $x^i$. Standard Finsler theories are
elaborated on tangent bundles with nontrivial N--connection structure (when $%
y^a$ are typical "fiber" coordinates which can be identified with certain
"velocity" fields if sections on basic manifolds are considered). In
general, Finsler like variables can be introduced on arbitrary manifolds, or
bundle spaces (in particular, parametrizing exact solutions in Einstein
gravity), see detailed discussions and examples in Refs. \cite%
{vrflg,vqgrbr,vgensol,ijgmmp}. In GR, the (Finsler like) N--connection
coefficients parametrize certain classes of nonholonomic frames and
off--diagonal metrics when coordinates $y^a$ are considered for a 2+2
splitting (but not as some "velocity" type variables). The concept of
nonolonomic manifold allows us to formulate an unified geometric approach
for all classes of such geometric and physical models.

\subsubsection{Einstein--Finsler spaces}

An Einstein manifold (space) is defined in standard form by a Levi--Civita
connection $\nabla =\{\ _{\shortmid }\Gamma _{\ \alpha \beta }^{\gamma } \} $
satisfying the field equations $\ _{\shortmid }R_{ij}=\lambda g_{ij},$ for a
(pseudo) Riemannian manifold $M$ endowed with metric $g_{ij},$ where $%
\lambda $ is the cosmological constant. In order to apply the anholonomic
frame method for constructing exact solutions from higher dimension gravity
\cite{vgensol}, we should introduce ''three shells of anisotropy''.\footnote{%
Parametrization I with ''anisotropic shells'' for higher order anisotropic
extensions, see details in Ref. \cite{vgensol}, when $T\mathbf{V}=hT\mathbf{V%
}\oplus \ ^{0}vT\mathbf{V}\oplus \ ^{1}vT\mathbf{V\oplus \ }^{2}vT\mathbf{%
\mathbf{V},}$ for local coordinates $\ u^{\ ^{1}\alpha }=(u^{\ ^{0}\alpha
},u^{\ ^{1}a})=(u^{i},u^{\ ^{0}a},u^{\ ^{1}a}),\ u^{\ ^{2}\alpha }=(u^{\
^{1}\alpha },u^{\ ^{2}a})=(u^{i},u^{\ ^{0}a},u^{\ ^{1}a},u^{\ ^{2}a}),$ with
$i,j,...=1,2,...,n;$ $\ ^{0}a,\ ^{0}b,...=n+1,...,n+m;$ $\ ^{1}a,\
^{1}b,...=n+m+1,...,n+m+...+\ ^{1}m;\ ^{2}a,\ ^{2}b,...=n+m+1,...,n+m+...+\
^{1}m,n+m+...+\ ^{1}m+...+\ ^{2}m$. We can consider $n=4$ and $m=\ ^{1}m=\
^{2}m=2,$ when $\dim M=2,$ or $4,\dim \mathbf{V=8}.$ Parametrization II is
for $n=4,m=4,\ ^{1}m=0,$ with trivial ''shall'' $\left( u^{\ ^{1}a}\right)$
and local coordinates $u^\alpha= (x^i,y^a),$ for $i=1,2,3,4$ and
consequently $a=5,6,7,8$ when on tangent bundles $5$ can be contracted to $%
1, $ $6$ to $2$ and so on. Parametrization III is for $n=2$ and $m=2,$ when $%
\dim M=2,\dim \mathbf{V}=4, $ with indices $i=1,2$ and $a=3,4$.} For a
d--connection $\mathbf{D}$ which is metric compatible with $\mathbf{g}$ (\ref%
{gdm}) on $\mathbf{V},$ we can consider generalized Einstein spaces defined
by
\begin{eqnarray}
R_{ij} =\ ^{h}\lambda (u)g_{ij},\ R_{\ ^{0}a\ ^{0}b} & =& \ ^{^{0}v}\lambda
(u)h_{\ ^{0}a\ ^{0}b},  \notag \\
R_{\ ^{1}a\ ^{1}b} =\ ^{^{1}v}\lambda (u)h_{\ ^{1}a\ ^{1}b},\ R_{\ ^{2}a\
^{2}b} &=&\ ^{^{2}v}\lambda (u)h_{\ ^{2}a\ ^{2}b},  \notag \\
R_{a_{0}i}=R_{i\ ^{0}a} =0, R_{\ ^{2}ai}=R_{i\ ^{2}a}=0,\ R_{\ ^{1}ai}=R_{i\
^{1}a} &=&0,  \notag \\
R_{\ ^{2}a\ ^{1}a}=R_{\ ^{1}a\ ^{2}a}=0,\ R_{\ ^{2}a\ ^{0}a}=R_{\ ^{0}a\
^{2}a} &=&0,  \label{esa1}
\end{eqnarray}%
where $\ ^{h}\lambda (u)$ and $\ ^{\ ^{0}v}\lambda (u),\ ^{\ ^{1}v}\lambda
(u),\ ^{\ ^{2}v}\lambda (u)$ are respectively the so--called locally
anisotropic $h$-- and $\ ^{0}v$--, $^{1}v$--, $^{2}v$--polarized
gravitational ''constants''. Such polarizations should be defined for
certain well defined constraints on matter and gravitational field dynamics,
lifts on tangent bundles, corrections from quantum gravity or any extra
dimension gravitational theory.

In this work, an Einstein--Finsler space is defined by a triple $\left[
\mathbf{N},\mathbf{g,D}\right] $ with a metric compatible d--connection $%
\mathbf{D}$ subjected to the condition to be a solution of equations (\ref%
{esa1}) with sources of type $\mathbf{\Upsilon }_{\ \ \ ^{3}\beta }^{\
^{3}\alpha }=diag[ _{\ ^{3}\gamma}\mathbf{\Upsilon }].$ Such equations can
be solved in very general form for the canonical d--connection $\widehat{%
\mathbf{D}}$ and certain nonholonomic restrictions to the Levi--Civita
connection $\nabla ,$ see \cite{vgensol,vsgg,vstr2}. In a general context,
we can consider that an Einstein--Finsler space is determined by a set of
solutions of (\ref{ensteqcdc}) with given sources and for a triple $\left[
\mathbf{N},\mathbf{g,D}\right] $ when $\mathbf{D}=\widehat{\mathbf{D}},$ or $%
\ ^{c}\mathbf{\mathbf{D}}$.

\subsubsection{Metric ansatz and partial differential equations}

\label{sseinst}

\paragraph{Third order anisotropic ansatz:}

\ Any metric (\ref{gdm}) can be reparameterized in a form with three shell
anisotropy, {\small
\begin{eqnarray}
\ \ \mathbf{g} &=&\ g_{ij}(x)dx^{i}\otimes dx^{j}+h_{\ ^{0}a\ ^{0}b}(x,\
^{0}y)\mathbf{e}^{\ ^{0}a}\otimes \mathbf{e}^{\ ^{0}b} +  \label{2forman} \\
&&h_{\ ^{1}a_{1}\ ^{1}b}(x,\ ^{0}y,\ ^{1}y)\mathbf{e}^{\ ^{1}a}\otimes
\mathbf{e}^{\ ^{1}b}+h_{\ ^{2}a\ ^{2}b}(x,\ ^{0}y,\ ^{1}y,\ ^{2}y)\mathbf{e}%
^{\ ^{2}a}\otimes \mathbf{e}^{\ ^{2}b},  \notag \\
\ \mathbf{e}^{\ ^{0}a} &=&dy^{\ ^{0}a}+N_{i}^{\ ^{0}a}(\ ^{0}u)dx^{i},\
\mathbf{e}^{\ ^{1}a}= dy^{\ ^{1}a}+N_{i}^{\ ^{1}a}(\ ^{1}u)dx^{i}+N_{\
^{0}a}^{\ ^{1}a}(\ ^{1}u)\ \mathbf{e}^{\ ^{0}a},  \notag \\
\mathbf{e}^{\ ^{2}a} &=&dy^{\ ^{2}a}+N_{i}^{\ ^{2}a}(\ ^{2}u)dx^{i}+N_{\
^{0}a}^{\ ^{0}a}(\ ^{2}u)\ \mathbf{e}^{\ ^{0}a}+N_{\ ^{1}a}^{\ ^{2}a}(\
^{2}u)\ \mathbf{e}^{\ ^{1}a},  \notag
\end{eqnarray}%
} for $x=\{x^{i}\},\ ^{0}y=\{y^{\ ^{0}a}\},\ ^{1}y=\{y^{\ ^{1}a}\},\
^{2}y=\{y^{\ ^{2}a}\},$ when the vertical indices and coordinates split in
the form $y=[\ ^{0}y,\ ^{1}y,\ ^{2}y],$ or $y^{a}=[y^{\ ^{0}a},y^{\
^{1}a},y^{\ ^{2}a}];$ $\ ^{0}u=(x,\ ^{0}y),\ ^{1}u=(\ ^{0}u,\ ^{1}y),\
^{2}u=(\ ^{1}u,\ ^{2}y),$ or $u^{\ ^{0}\alpha }=(x^{i},y^{\ ^{0}a}),u^{\
^{1}\alpha }=(u^{\ ^{0}\alpha },y^{\ ^{1}a}),u^{\ ^{2}\alpha }=(u^{\
^{1}\alpha },y^{\ ^{2}a}).$

There is a very general ansatz of this form (with Killing symmetry on $%
y^{8}, $ when the metric coefficients do not depend on variable $y^{8};$ it
is convenient to write $\ y^{3}=\ ^{0}v,$ $y^{5}=\ ^{1}v,y^{7}=\ ^{2}v$ and
introduce parametrization of $N$--coefficients via $n$-- and $w$--functions)
defining exact solutions of (\ref{esa1}), {\small
\begin{eqnarray}
\ ^{sol}\mathbf{g} &=&g_{i}(x^{k}){dx^{i}\otimes dx^{i}}+h_{\ ^{0}a}(x^{k},\
^{0}v)\mathbf{e}^{\ ^{0}a}{\otimes }\mathbf{e}^{\ ^{0}a}  \label{ansgensol}
\\
&&+h_{\ ^{1}a}(u^{\ ^{0}\alpha },\ ^{1}v)\ \mathbf{e}^{\ ^{1}a}{\otimes }\
\mathbf{e}^{\ ^{1}a} +h_{\ ^{2}a}(u^{\ ^{1}\alpha },\ ^{2}v)\ \mathbf{e}^{\
^{2}a}{\otimes }\ \mathbf{e}^{\ ^{2}a},  \notag \\
\mathbf{e}^{3} &=&dy^{3}+w_{i}(x^{k},\ ^{0}v)dx^{i},\mathbf{e}%
^{4}=dy^{4}+n_{i}(x^{k},\ ^{0}v)dx^{i},  \notag \\
\mathbf{e}^{5} &=&dy^{5}+w_{\ ^{0}\beta }(u^{\ ^{0}\alpha },\
^{1}v)du^{\beta },\mathbf{e}^{6}=dy^{6}+n_{\ ^{0}\beta }(u^{\ ^{0}\alpha },\
^{1}v)du^{\ ^{0}\beta },  \notag \\
\mathbf{e}^{7} &=&dy^{7}+w_{\ ^{1}\beta }(u^{\ ^{1}\alpha },\ ^{2}v)du^{\
^{1}\beta },\mathbf{e}^{8}=dy^{8}+n_{\ ^{1}\beta }(u^{\ ^{1}\alpha },\
^{2}v)du^{\ ^{1}\beta }.  \notag
\end{eqnarray}
} In Theorem 1.1 of \cite{vgensol} (we should consider those results for
three shells and trivial $\omega $--coefficients), there are stated explicit
conditions on $w$-- and $n$--coefficients and $\mathbf{\Upsilon }$--sources,
for arbitrary dimensions and in very general forms, when an ansatz (\ref%
{ansgensol}) generates exact solutions gravity.

\paragraph{Separation of equations for the canonical d--connection:}

Let us consider the ansatz (\ref{ansgensol}) for dimensions $n=2$ and $\
^{0}m=2,\ ^{1}m=$ $\ ^{2}m=0,$ when the source is parameterized in the form $%
\Upsilon _{\ \ \beta }^{\alpha }=diag[\Upsilon _{\gamma };\Upsilon
_{1}=\Upsilon _{2}=\Upsilon _{2}(x^{k});\Upsilon _{3}=\Upsilon _{4}=\Upsilon
_{4}(x^{k},y^{3})]$. Computing the corresponding coefficients of
d--connection $\widehat{\mathbf{D}}$ following formulas (\ref{candcon}) and
introducing them respectively into (\ref{dcurv}) (\ref{dricci}), we express
the gravitational field equations (\ref{ensteqcdc}) in the form\footnote{%
the details of such computations can be found in Part II of \cite{vsgg} and
in \cite{vgensol}} {\small
\begin{eqnarray}
&& \widehat{R}_{1}^{1} =\widehat{R}_{2}^{2} =-\frac{1}{2g_{1}g_{2}}%
[g_{2}^{\bullet \bullet }-\frac{g_{1}^{\bullet }g_{2}^{\bullet }}{2g_{1}}-%
\frac{\left( g_{2}^{\bullet }\right) ^{2}}{2g_{2}}+g_{1}^{\prime \prime }-%
\frac{g_{1}^{\prime }g_{2}^{\prime }}{2g_{2}}-\frac{\left( g_{1}^{\prime
}\right) ^{2}}{2g_{1}}] =-\Upsilon _{2}(x^{k}),  \label{eqe1} \\
&&\widehat{R}_{3}^{3} =\widehat{R}_{4}^{4}=-\frac{1}{2h_{3}h_{4}}[
h_{4}^{\ast \ast }-\frac{\left( h_{4}^{\ast }\right) ^{2}}{2h_{4}}-\frac{%
h_{3}^{\ast }h_{4}^{\ast }}{2h_{3}}] =-\Upsilon _{4}(x^{k},y^{3}),
\label{eqe2} \\
&&\widehat{R}_{3k} =\frac{w_{k}}{2h_{4}}[h_{4}^{\ast \ast}-\frac{%
(h_{4}^{\ast })^{2}}{2h_{4}}-\frac{h_{3}^{\ast}h_{4}^{\ast}}{2h_{3}}] +\frac{%
h_{4}^{\ast }}{4h_{4}}(\frac{\partial _{k}h_{3}}{h_{3}}+\frac{\partial
_{k}h_{4}}{h_{4}})-\frac{\partial _{k}h_{4}^{\ast }}{2h_{4}}=0  \label{eqe3a}
\\
&&\widehat{R}_{4k}=\frac{h_{4}}{2h_{3}}n_{k}^{\ast \ast }+\left( \frac{h_{4}%
}{h_{3}}h_{3}^{\ast }-\frac{3}{2}h_{4}^{\ast }\right) \frac{n_{k}^{\ast }}{%
2h_{3}}=0.  \label{eqe3}
\end{eqnarray}%
} In the above formulas, we denote $a^{\bullet }=\partial a/\partial x^{1},$%
\ $a^{\prime }=\partial a/\partial x^{2},$\ $a^{\ast }=\partial a/\partial
y^{3}.$

The system (\ref{eqe1})--(\ref{eqe3}) is nonlinear and with partial
derivatives. Nevertheless, the existing separation of equations (we should
not confuse with separation of variables which is a different property)
allows us to construct very general classes of exact solutions (depending on
conditions if certain partial derivatives are zero, or not). For any
prescribed $\Upsilon _{2}(x^{k}),$ we can define $g_{1}(x^{k})$ (or,
inversely, $g_{2}(x^{k})$) for a given $g_{2}(x^{k})$ (or, inversely, $%
g_{1}(x^{k})$) as an explicit, or non--explicit, solution of (\ref{eqe1}) by
integrating two times on $h$--variables. Similarly, taking any $\Upsilon
_{4}(x^{k},y^{3}),$ we solve (\ref{eqe2}) by integrating one time on $y^{3}$
and defining $h_{3}(x^{k},y^{3})$ for a given $h_{4}(x^{k},y^{3})$ (or,
inversely, by integrating two times on $y^{3} $ and defining $%
h_{4}(x^{k},y^{3})$ for a given $h_{3}(x^{k},y^{3})$).

Haven determined the values $g_{i}(x^{k})$ and $h_{\ ^{0}a}(x^{k},y^{3}),$
we can compute the coefficients of N--connection:\ The functions $%
w_{j}(x^{k},y^{3})$ are solutions of algebraic equations (\ref{eqe3a}).
Finally, we have to integrate two times on $y^{3}$ in order to obtain $%
n_{j}(x^{k},y^{3}).$ Such general solutions depend on integration functions
depending on coordinates $x^{k}.$ In physical constructions, we have to
consider well defined boundary conditions for such integration functions.

\paragraph{Equations for the h-v / Cartan d--connection:}

\ In N--adapted frames, the h--v d--connection $\widetilde{\mathbf{D}}$ is
determined by coefficients $\widetilde{\mathbf{\Gamma }}_{\ \beta \gamma
}^{\alpha }=\left( \widetilde{L}_{\ bk}^{a},\widetilde{C}_{bc}^{a}\right) ,$
{\small
\begin{equation}
\widetilde{L}_{\ jk}^{i} =\frac{1}{2}g^{ih}(\mathbf{e}_{k}g_{jh}+\mathbf{e}%
_{j}g_{kh}-\mathbf{e}_{h}g_{jk}),\ \widetilde{C}_{\ bc}^{a} =\frac{1}{2}%
h^{ae}(e_{b}h_{ec}+e_{c}h_{eb}-e_{e}h_{bc}),  \label{cdc}
\end{equation}
} are computed for a d--metric $\mathbf{g=}[g_{ij},h_{ab}]$ (\ref{gdm}). Via
frame transforms to $\ ^{F}\mathbf{g}$ (\ref{fsm}), $\mathbf{g}_{\alpha
^{\prime }\beta ^{\prime }}=e_{\ \alpha ^{\prime }}^{\alpha }e_{\ \beta
^{\prime }}^{\beta }\ \ ^{F}\mathbf{g}_{\alpha \beta },$\ we can define such
N--adapted frames when the coefficients of $\ \widetilde{\mathbf{D}}$ are
equal to the coefficients of the Cartan d--connection $\ ^{c}\mathbf{D}=\
\{^{c}L_{ik}^{i},\ ^{c}C_{bc}^{a}\}$ (\ref{cartand}). For dimensions $n=2$
and $\ ^{0}m=2,\ ^{1}m=$ $\ ^{2}m=0,$ the equations (\ref{esa1}) for $%
\widetilde{\mathbf{\Gamma }}_{\ \alpha \ \beta }^{\ \gamma }$ (\ref{cdc})
and source, transform into an exactly integrable system of partial
differential equations when the coefficients of (\ref{gdm}) are stated by an
ansatz $\mathbf{g}_{\alpha \beta }=diag[g_{i}(x^{k}),h_{a}(x^{i},v)]$ and $%
N_{k}^{3}=w_{k}(x^{i},v),N_{k}^{4}=n_{k}(x^{i},v).$ The first two equations
are equivalent, respectively, to (\ref{eqe1}) and (\ref{eqe2}), \ {\small
\begin{eqnarray}
\widetilde{R}_{1}^{1} &=&\widetilde{R}_{2}^{2}=\widehat{R}_{1}^{1}=\widehat{R%
}_{2}^{2}=-\ \Upsilon _{2}(x^{k}),  \label{ep1a} \\
\widetilde{R}_{3}^{3} &=&\widetilde{R}_{4}^{4}=\widehat{R}_{3}^{3}=\widehat{R%
}_{4}^{4}=\ -\ \Upsilon _{4}(x^{k},y^{3}).  \label{ep2a}
\end{eqnarray}%
} Instead of (\ref{eqe3a}), (\ref{eqe3}) we get, correspondingly, {\small
\begin{eqnarray}
\ \widetilde{R}_{3j} &=&\frac{h_{3}^{\ast }}{2h_{3}}w_{j}^{\ast }+A^{\ast
}w_{j}+B_{j}=0,  \label{ep3a} \\
\widetilde{R}_{4i} &=&-\frac{h_{4}^{\ast }}{2h_{3}}n_{i}^{\ast }+\frac{%
h_{4}^{\ast }}{2}K_{i}=0,  \label{ep4a}
\end{eqnarray}
\begin{eqnarray}
\mbox{ where } A &=&\left( \frac{h_{3}^{\ast }}{2h_{3}}+\frac{h_{4}^{\ast }}{%
2h_{4}}\right) ,\ B_{k}=\frac{h_{4}^{\ast }}{2h_{4}}\left( \frac{\partial
_{k}g_{1}}{2g_{1}}-\frac{\partial _{k}g_{2}}{2g_{2}}\right) -\partial _{k}A,
\label{aux} \\
K_{1} &=&-\frac{1}{2}\left( \frac{g_{1}^{\prime }}{g_{2}h_{3}}+\frac{%
g_{2}^{\bullet }}{g_{2}h_{4}}\right) ,\ K_{2}=\frac{1}{2}\left( \frac{%
g_{2}^{\bullet }}{g_{1}h_{3}}-\frac{g_{2}^{\prime }}{g_{2}h_{4}}\right) .
\notag
\end{eqnarray}
}

The system (\ref{ep1a})--(\ref{ep4a}) also has the property of separation of
equations. In this case (having defined $h_{a}(x^{i},v)),$ we can compute $%
n_{k}(x^{i},v)$ integrating the equation (\ref{ep4a}) on $y^{3}=v$ and,
respectively, solving an usual first order differential equation (\ref{ep3a}%
), on $y^{3}=v,$ considering $x^{i}$ as parameters. Prescribing a generating
function $F(x^{i},v),$ such a solution given by data $%
g_{i}(x^{k}),h_{a}(x^{i},v),$ and $%
N_{k}^{3}=w_{k}(x^{i},v),N_{k}^{4}=n_{k}(x^{i},v)$ can be represented
equivalently as a (pseudo) Finsler space. To associate such a
h--v--configuration to a real Finsler geometry is convenient to work with
sets of local carts on $TM,$ or $\mathbf{V},$ when the quadratic algebraic
system for $e_{\ \alpha ^{\prime }}^{\alpha }$ has well defined real
solutions.

\paragraph{General solutions for Finsler gravity:}

\ The system of gravitational field equations with three shell anisotropy
(when its first shell restriction given by (\ref{eqe1})--(\ref{eqe3})) can
be solved in general form following the results of the mentioned above
Theorem 1.1 from \cite{vgensol}. We omit in this work such cumbersome
formulas (3d order solutions for anisotropic canonical d--connections are
considered in Appendix \ref{asect2}) but give an explicit example of
anisotropic generalization of Friedman--Robertson--Walker (FRW) solutions
depending on time and three velocity coordinates in subsection \ref{ssofdfa}.

In this subsection, we provide the general solution of equations (\ref{ep1a}%
)--(\ref{ep4a}) for the h--v/Cartan d--connection. It can be written for a
d--metric $\mathbf{g}=[g_{ij},h_{ab}]$ (\ref{gdm}) with coefficients
computed in the form {\small
\begin{eqnarray}
g_{i} &=&\epsilon _{i}e^{\psi (x^{k})},\mbox{\ for }\epsilon _{1}\psi
^{\bullet \bullet }+\epsilon _{2}\psi ^{\prime \prime }=\Upsilon _{2}(x^{k});
\label{sol1} \\
h_{3} &=&\epsilon _{3}\ ^{0}h(x^{i})\ [f^{\ast }(x^{i},v)]^{2}|\varsigma
(x^{i},v)|,  \notag \\
\mbox{for }\varsigma &=&\ ^{0}\varsigma (x^{i})-\frac{\epsilon _{3}}{8}\
^{0}h(x^{i})\int dv\ \ \Upsilon _{4}(x^{k},v) f^{\ast }(x^{i},v)\
[f(x^{i},v)-\ ^{0}f(x^{i})],  \notag \\
h_{4} &=&\epsilon _{4}[f(x^{i},v)-\ ^{0}f(x^{i})]^{2};  \notag \\
w_{j} &=&\ _{0}w_{j}(x^{i})\exp \{ -\int_{0}^{v}\left[ \frac{2h_{3}A^{\ast }%
}{h_{3}^{\ast }}\right] _{v\rightarrow v_{1}}dv_{1} \} \int_{0}^{v}dv_{1}%
\left[ \frac{h_{3}B_{j}}{h_{3}^{\ast }}\right] _{v\rightarrow v_{1}}  \notag
\\
&& \exp \{ -\int_{_{0}}^{v_{1}}\left[ \frac{2h_{3}A^{\ast }}{h_{3}^{\ast }}%
\right] _{v\rightarrow v_{1}}dv_{1} \},\ n_{i} = \ _{0}n_{i}(x^{k})+\int dv\
h_{3}K_{i}.  \notag
\end{eqnarray}%
} Such solutions with $h_{3}^{\ast }\neq 0$ and $h_{4}^{\ast }\neq 0$ are
determined by generating functions $f(x^{i},v),f^{\ast }\neq 0,$ and
integration functions $\ ^{0}f(x^{i}),\ ^{0}h(x^{i}),$ $\ _{0}w_{j}(x^{i})$
and $\ _{0}n_{i}(x^{k});$ the coefficients $A$ and $B_{j},K_{i}$ are given
by formulas (\ref{aux}).\footnote{%
we can prove that such coefficients generate general solutions by
straightforward computations being similar to those for the canonical
d--connections, see details in Refs. \cite{vsgg,vgensol}}

\subsection{Principles of general relativity and Finsler gravity}

The goal of this section is to show how the theoretic scheme and principles
for GR can be extended to metric compatible Finsler gravity theories.

At present, all classical gravitational phenomena are completely described
the standard GR (one may be some exceptions for dark energy and dark matter
theories; we shall discuss the problem in section \ref{smfctb}). However, it
is generally accepted that the incompatibility between GR and quantum theory
which should be treated in a more complete theory of quantum gravity (QG)
which is under elaboration. It is supposed also that from any general theory
(string/brane gravity, commutative and noncommutative gauge or other
generalizations, various quantum models etc) GR is reproduced with small
corrections in the classical limit. Such small corrections necessarily
violate the \textbf{equivalence principle} in GR \cite{colladay}. This also
results in violation of local \textbf{Lorenz invariance} both in special
relativity (SR) and GR (we explained in section \ref{ssvls} how nonlinear
dispersion relations from any QG model result in Finsler type metrics
depending on velocitiy/momentum variables).

\subsubsection{Minimal extensions of Einstein gravity to Finsler theories}

Different researches in geometry and physics work with different concepts of
Finsler space. In order to avoid ambiguities, let us state what rules we
follow in this paper in order to elaborate a class of Finsler gravity
theories which seem to be most closed to the modern paradigm of standard
physics.

A (pseudo) Finsler geometry model is canonically defined by a (fundamental
Finsler function $F(x,y)$ on a N--anholonomic manifold $\mathbf{V}$ (in
particular, on a tangent bundle $TM$) with prescribed h--/ v--splitting.
Such a model is completely defined on $T\mathbf{V}$ if there are formulated
certain geometric principles for generating by $F,$ in a unique form, a
triple of fundamental geometric objects $\left( \ ^{F}\mathbf{N,\ ^{F}g,\
^{F}D}\right)$. Via frame transforms, any such a triple transforms into a
general one $\left( \mathbf{N,g,D}\right),$ with a decompositions $\nabla =%
\mathbf{D}+\mathbf{Z},$ when a d--connection $\mathbf{D,}$ a Levi--Civita
connection $\nabla $ and a distorting tensor $\mathbf{Z}$ are uniquely
determined by $\mathbf{N}$ and $\mathbf{g.}$ Inversely, we can always
introduce Finsler variables (with left up $F$--label) for any data $\left(
\mathbf{N,g,D}\right).$ For simplicity, we shall omit hereafter the label $F
$ is that will not result in ambiguities.

There are two general classes of Finsler geometries: 1) metric compatible,
which can be included into (or minimally extending) standard theories, when $%
\mathbf{\ Dg=0,}$ and 2) metric noncompatible (generating nonstandard
models), when $\mathbf{Dg=Q\neq 0.}$ For any given $F,$ and/or $\mathbf{g,}$
there is a unique canonical d--connection $\widehat{\mathbf{D}}$ (\ref%
{candcon}) when $\widehat{\mathbf{D}}\mathbf{g}=0$ and $h$- and $v$%
--torsions vanish but the general nonzero torsion $\widehat{\mathcal{T}}$ is
induced by $\mathbf{N}$ and $\mathbf{g}.$ On spaces of odd dimension, $%
\widehat{\mathbf{D}}$ can be transformed into the h--v d--connection $%
\widetilde{\mathbf{D}}$ is determined by a couple of coefficients $%
\widetilde{\mathbf{\Gamma }}_{\ \beta \gamma }^{\alpha }=\left( \widetilde{L}%
_{\ bk}^{a},\widetilde{C}_{bc}^{a}\right) $ (\ref{cdc}). Via frame
transforms, we can relate this d--connection to the Cartan d--connection $\
^{c}\mathbf{D}=\ \{^{c}L_{ik}^{i},\ ^{c}C_{bc}^{a}\}$ (\ref{cartand}) for
Finsler geometry, which also defines a canonical almost symplectic
d--connection, see details in \cite{matsumoto,ma,vrflg,vqgrbr}. We can work
equivalently on a nonholonomic space $\mathbf{V}$ with any $\widehat{\mathbf{%
D}},\widetilde{\mathbf{D}}$ and/or $\ ^{c}\mathbf{D}$ in N--adapted form.

The key issues for elaborating a Finsler generalization of Einstein gravity
are to introduce on $\mathbf{V}$ (in particular on $TM$) a metric $\mathbf{g}
$ with pseudo--Euclidean signature and to decide what type of metric
compatible d--connec\-ti\-on $\mathbf{D}$ will be used for postulating the
field (Einstein--Finsler) equations. For physically viable Finsler gravity
theories, any generalized Finsler fundamental geometric objects $g$ and $D$
should contain as particular cases certain Einstein gravity solutions. We
proved some important results \cite{vgensol,ijgmmp,vsgg,vrflg} that for $%
\widehat{\mathbf{D}}$ and/or $\widetilde{\mathbf{D}}\longleftrightarrow \
^{c}\mathbf{D,}$ the Einstein equations can be integrated in very general
forms. Imposing the \textquotedblright zero torsion\textquotedblright\
constraints (\ref{lccond}), when $\widehat{\mathbf{D}}\rightarrow \nabla ,$
we restrict the integral varieties to define general solutions in Einstein
gravity and its higher dimension generalizations.

\subsubsection{Principles of Einstein--Finsler relativity}

\paragraph{Finsler configurations in general relativity:}

\ Finsler variables and the canonical d--connection $\widehat{\mathbf{D}}%
=\nabla -\widehat{\mathbf{Z}}$ \ (\ref{distorsrel}), (similarly, $\widetilde{%
\mathbf{D}}$ and/or $\ ^{c}\mathbf{D)}$ can be introduced in general
relativity if nonholonomic 2+2 splitting are considered for a nonholonomic
pseudo--Riemannian spacetime $\mathbf{V}$ \cite{vrflg,vqgrbr}. All geometric
and physical objects and fundamental equations can be re--expressed in terms
of $\widehat{\mathbf{D}}$ and N--adapted variables. Such a formal Finsler
gravity satisfies all axioms introduced for the Einstein gravity theory. So,
alternatively to well known tetradic, spinor, Ashtekar and other variables
in general relativity, we can introduce nonholonomic/Finsler variables.

\paragraph{Minimal Finsler extensions on $TM$ of the standard model:}

The concept of flat Min\-kow\-ski spacetime, with pseudo--Euclidean
signatures, and postulates for SR are related to the Maxwell electromagnetic
field theory. They where formulated following Michelson--Morley type
experiments with constant speed of light. The most important symmetry is
that of Lorentz (pseudo--rotation) and Poincare (with translations)
invariance with respect to certain linear group transforms.

Any possible contributions from QG result in nonlinear dispersions for light
rays of type (\ref{disp}) and nonlinear quadratic elements (\ref{fbm}). In
order to explain such physical effects and elaborate generalized models of
classical and quantum theories there were considered various
generalizations/restrinctions of symmetries in SR \cite%
{girelli,gibbons,stavr1,mignemi} when the Minkowski metric $\eta
_{ij}=[-1,1,1,1]$ transform into a Finsler type $\mathbf{g}_{ij}(y)$
depending locally on velocity/momentum type coordinates $y^{i}.$ It is not
clear from general physical arguments why certain models of broken Lorentz
invariance should have priorities with respect to another ones. Perhaps, in
a ''minimal way'' we can say that $\eta _{ij}\rightarrow g_{ij}(y)$ is
similar in some lines with generalizations of SR to GR, $\eta
_{ij}\rightarrow g_{ij}(x),$ but in our case we may have an additional
curvature determined by fibers of a co/tangent bundle, in general, by
metrics of form $g_{ij}(x,y).$ Such a metric should be a solution of the
Einstein--Finsler equations and may possess some nonholonomically deformed
Lorentz symmetries.

\vskip2pt \textbf{Generalized equivalence principle:} In Newtonian theory of
gravity, the experimental data show that the gravitational force on a body
is proportional to its inertial mass. This supports a fundamental idea that
all bodies are influenced by gravity and, indeed, all bodies fall precisely
the same way in gravitational fields. \ Because motion is independent of the
nature of the bodies, the paths of freely falling bodies define a preferred
set of curves in spacetime just as in special relativity the paths in
spacetime of inertial bodies define a preferred set of curves.

The world lines of freely falling bodies in a gravitational field are simply
the geodesics of the (curved) spacetime metric. This suggests the
possibility of ascribing properties of the gravitational field to the
structure of spacetime itself. Because nonlinear dispersions from Minkowski
spacetime can be associated to metrics of type $g_{ij}(y),$ and GR to
metrics of type $g_{ij}(x),$ we can consider a generalized equivalence
principle on Finsler spacetimes with metrics of type $\ ^{F}g_{ij}(x,y).$ We
may preserve the ideas of Universality of Free Fall and the Universality of
the Gravitational Redshift in a Finsler type spacetime modelled by data $%
\left( \mathbf{N,g,D}\right) .$ \ In such a locally anisotropic spacetime,
the paths of freely falling bodies are not usual geodesics but certain
nonlinear (semi--spray) ones which are different from auto--parallels of $%
\mathbf{D,}$ see details on such geometries in \cite{vrflg,vsgg,ma,bcs} and
references therein.\footnote{%
For models of generalized Finsler spacetimes, it is important to study the
geometry of semi--spray configurations as N--connection generalizations of
autoparallel and geodesic curves. In some sense, semi--sprays characterize
the N--connection effects into "physical paths" of test particles. We
recommend the interested readers to consult the respective sections in the
mentioned review papers and monographs.}

Working with metric compatible d--connections completely determined by the
metric and N--connection structures, we can establish a 1--1 correspondence
between one type of preferred curves (semi--sprays) and auto--parallels of $%
\mathbf{D.}$ This way we can encode equivalently the experimental (curvature
deviation) data with respect to both types of congruences. \ In all
important physical equations for a Finsler gravitational and matter fields,
the connection $\mathbf{D}$ (for canonical constructions, it is used $\
\widehat{\mathbf{D}},$ $\widetilde{\mathbf{D}}$ and/or $\ ^{c}\mathbf{D}$)
is contained. Such a d--connection can be used for constructing the Dirac,
d'Alambert and other important operators which allows us to compute the
light and particle propagation in a Finsler spacetime.

\vskip2pt \textbf{Generalized Mach principle: } The Einstein gravity theory
was formulated using a second much less precise set of ideas which goes
under the name of Mach`s principle. In SR and in pre--relativity notions of
spacetime, the geometric structure of spacetime is given once and for all
and is unaffected by the material bodies that may be present. In particular,
the properties of ``inertial motion`` and ``non--rotating`` are not
influenced by matter in the universe. Mach supposed that all matter in the
universe should contribute to the local definition of ''non--acceleration''
and ''non--rotating''. Einstein accepted this idea and was strongly
motivated to formulate a theory where, unlike SR, the structure of spacetime
is influenced by the presence of matter. In GR, such purposes were achieved
only partially. With respect to Finsler gravity theories on (co) tangent
bundles derived from quantum nonlinear dispersion we can consider a
generalized Mach principle that quantum energly/motion should contribute to
spacetime, i.e. the structure of spacetime is influenced by the presence of
quantum world. This influence is encoded both into the nonholonomic
structure and via coefficients of $\left( \mathbf{N,g,D}\right) $ into
energy--momentum tensors for matter fields imbedded self--consistently in
spacetime aether with moving coordinates $y^{a}.$

\vskip2pt \textbf{Einstein--Finsler spacetimes and gravitational equations: }
New theories of locally anisotropic spacetime and gravitation state the
following:\ The intrinsic, observer--independent, properties of Finsler
spacetime are described by a Finsler generating functions which canonically
determine the N--connection, d--metric and d--connection fundamental
geometric objects in a metric compatible form as in GR but for N--adapted
constructions. We define a Finsler gravity model and its fundamental
gravitational equations on a N--anholonomic manifold $\mathbf{V},$ including
GR and SR (as certain particular classes of solutions) following the same
principles as in Einstein theory but in N--adapted form for a fixed
canonical metric compatible d--connection (a Finsler d--connection) which is
different from the Levi--Civita connection.

\vskip2pt \textbf{Principle of general covariance:} In GR, this is a natural
consequence from that fact that spacetime models are constructed on (pseudo)
Riemannian manifolds. So, the geometric and physical constructions do not
depend on frames of reference (observers) and coordinate transforms. In
definition of Finsler geometry models the concept of manifold is also
involved (in certain approaches such manifolds are tangent/vector bundle
spaces). So, the principles of general covariance has to be extended on $%
\mathbf{V,}$ or $TM.$ We can introduce certain preferred systems of
reference and adapted coordinate transforms when a fixed $h$--$v$%
--decomposition is preserved/distinguished but this is a property of some
particular classes of solutions of the Einstein--Finsler equations. In
general, \ we can not distinguish between triples of data $\left( \ ^{F}%
\mathbf{N,\ ^{F}g,\ ^{F}D}\right) $ and $\left( \mathbf{N,g,D}\right).$ We
can use any parametrizations of Finsler data which are necessary for certain
construction in a model of classical or quantum gravity. In an extended for
Finsler spaces principle of generalized covariance (for instance, for the
canonical d--connection), there are included distortion relations of type $%
\widehat{\mathbf{D}}=\nabla -\widehat{\mathbf{Z}}$ \ (\ref{distorsrel}). So,
we can describe geometrical and physical models equivalently both in terms
of $\ \widehat{\mathbf{D}}$ and $\nabla $ because such connections are
defined by the same metric structure.

\vskip2pt \textbf{The equations of motion and conservation laws: }The
conservation law $\nabla _{i}T^{ij}=0$ is a consequence of the Bianchi
relations and involve the idea that in GR the Einstein's equations alone
actually implies the geodesic hypothesis (that the world lines of test
bodies are geodesics of the spacetime metric). Note however, that bodies
which are ''large'' enough to feel the tidal forces of the gravitational
field will deviate from geodesic motion. \ Such deviations may be caused by
certain nonholonomic constraints on the dynamics of gravitational fields.
The equations of motion of such bodies in GR also can be found from the
condition $\nabla _{a}T^{ab}=0.$

For a Finsler d--connection $\mathbf{D,}$ even it is metric compatible, $%
\mathbf{D}_{\alpha }\Upsilon ^{\alpha \beta }\neq 0,$ which is a consequence
of non--symmetry of the Ricci and Einstein d--tensors, see explanations for
formula (\ref{enstdt}) and generalized Bianchi identities. Such a property
is also related to nonholonomic constraints on the dynamics of Finsler
gravitational fields. It is not surprising that the ''covariant divergence''
of source does not vanish even for $\ \widehat{\mathbf{D}},$ $\widetilde{%
\mathbf{D}}$ and/or $\ ^{c}\mathbf{D.}$ Using distorting relations of type $%
\widehat{\mathbf{D}}=\nabla -\widehat{\mathbf{Z}}$ \ (\ref{distorsrel}), we
can always compute $\widehat{\mathbf{D}}_{\alpha }\Upsilon ^{\alpha \beta }$
from $\nabla _{i}T^{ij}$ for matter fields moving in a canonical Finsler
spacetime following principles minimally generalizing those for general
relativity as we explained above. In this case, the conservation law are
more sophisticate by nonholonomic constraint but nevertheless it is possible
to compute effective nonholonomic tidal forces of locally anisotropic
gravitational fields when auto-parallels of $\widehat{\mathbf{D}}$ deviated
from nonlinear geodesic (semi--spray) configurations.

\vskip2pt \textbf{Axiomatics for the Einstein--Finsler gravity: \ } A
constructive--axiomatic approach to GR was proposed in 1964 by J. Ehlers, F.
A. E. Pirani and A. Schild \cite{eps} (the so--called EPS axioms). In a
series of publications in the early 1970's and further developments, see
original results and references in \cite{pir,wood,perlick,meist1}, it was
elaborated the concept of EPS spacetime as a physically motivated geometric
model of spacetime geometry. That axiomatic approach led to a common belief
that the underlying geometry of the spacetime can be only pseudo--Riemannian
which lead to the paradigmatic concept of ''Lorentzian 4--manifold'' in GR.

An axiomatic approach to Finsler gravity theory was proposed in \cite{pim2};
it was formulated also a minimal set of axioms for Finsler geometry \cite%
{matsumoto}. We consider that it is not possible to elaborate a general EPS
type system for all types of Finsler gravity theories. For the
Einstein--Finsler spaces, the EPS axioms can be extended on $TM$ when $%
\widetilde{\mathbf{D}}/\ ^{c}\mathbf{D}$ are used for definition of
auto--parallels and light propagation on nonholonomic manifolds.

\subsubsection{On physical meaning of velocity/momentum coordinates}

\label{ssphint}We can consider the GR theory as a branch of modern
mathematics when the physical theories are geometrized on a Lorentz manifold
$\mathbf{V.}$ The concepts of tangent Lorentz bundle, $T\mathbf{V,}$ and its
dual, $T^{\ast }\mathbf{V,}$ are well--defined. For corresponding local
coordinates $(x,y)=(x^{i},y^{a})$ and $(x,p)=(x^{i},p_{a}),$ the values $%
y^{a}$ are called "velocity" type coordinates and $p_{a}$ are "dual", or
"momentum" type coordinates. Geometrically, the gravity and cosmological and
physical models on $T\mathbf{V,}$ or $T^{\ast }\mathbf{V,}$ can be
elaborated as higher dimension ones for metric--affine, \cite{vsgg}, or
(super) Finsler superstring theories \cite{vstr2}, with "velocity/momentum"
extra--dimensional coordinates. Such theories can not be compactified on $%
y^{a}$ because there are  a constant velocity of light in vacuum and, in
general, a nontrivial N--connection structure. There are important various
off--diagonal trapping/warping effects for Finsler branes \cite%
{vbranesf,vhfl}. The corresponding metrics are generic off--diagonal, the
N-- and d--connections are with nontrival torsion, or/and with nonzero
nonmetricity fields, depending on velocity/momentum type variables. Using
N--adapted lifts and corresponding covariant differential and variational
calculus, we can extend geometrically any physical theory on $\mathbf{V}$ to
$T\mathbf{V}$ and/or $T^{\ast }\mathbf{V}$. We can consider such theories as
a simple framework for encoding properties of flat quantum space--time.

Nevertheless, only a formal geometric approach does not allow us to provide
a well--defined physical meaning to $y^{a}$ and/or $p_{a}.$ Such values can
be naturally introduced in geometric mechanics and effective (commutative
and noncommutative, or (super) symmetric) theories of gravity \cite{ma,vsgg}%
, but their physical interpretation depends on explicit form of models we
elaborate. Let us analyze and discuss in brief several important works developing
theories with $y^{a}$ and/or $p_{a}.$

\paragraph{Nonlinear deformed relativity (or doubly special relativity, DSR;
see \protect\cite{amelino,magueijo,kimberly} and references therein):}

This class of theories is constructed via non-linear realizations of the
Lorentz group when modified Lorentz transformations reduce to the usual ones
at low energies. There are considered deformed dispersion relations and an
invariant $E_{P}$ as a borderline between classical and quantum gravity. For
a number of historical and phenomenological reasons (the theory is motivated
by cosmic ray kinematics) the non-linear relativity was first studied on
momentum space when recovering the position space is highly non-trivial.

Considering a dual to non-linear realizations of relativity in momentum
space, there were constructed energy-dependent metrics, connection and
curvature for a simple modification to Einstein's equations. Several
counterparts to the cosmological metrics where found and shown how
cosmologies based upon "energy--depending" theory of gravity may solve the
horizon problem. There were explored some solutions to this theory of
gravity, namely the cosmological, black hole, and weak field solutions. Such
results may have important implications for black hole thermodynamics.

It should be noted that different theories of gravity are derived for
different realizations of position space. If the Lorentz group is
non-linearly realized in the position space, gravity is induced gauging a
symmetry which is non-linearly realized. \ It is possible to consider
noncommutative positions and/or momentum spaces.

\paragraph{Quantum-gravitational fluctuations in the space-time and \newline
D--brane/--particle foam (see \protect\cite%
{ellis,mavromatos,mavromatos1,mavromatos2}):}

The main idea was that quantum-gravitational fluctuations in the space-time
background induce non-trivial optical properties of the vacuum. That include
diffusion and stochastic properties and consequent uncertainties in the
arrival times of photons. Such an approach can be motivated within a Liouville
string formulation of quantum gravity and suggests a frequency--dependent
refractive index of the vacuum for particles and quantum fields. An explicit
realization can be constructed by treating photon propagation through
quantum excitations and, in general, anisotropic D-brane fluctuations in the
space-time foam. This way we can describe string effects that lead to
stochastic fluctuations in couplings and hence in the velocity of light.

Within the context of supersymmetric space--time (D--particle) foam in
string/brane--theory, it is possible to construct models of Finsler--induced
cosmology and to study possible implications for (thermal) dark matter
abundances. In such approaches, there are elaborated microscopic models of
dynamical space-time, where Finsler geometries arise naturally. For
instance, we can consider effects of recoil of D--particles related to a
back--reaction on the space-time metric of Fisler type which is
stochastic.The induced Finsler-type metric distortions depend (additionally
to the space-time coordinates) also on the pertinent momentum transfer which
provies an explicit physical interpretation of such coordinates.

\paragraph{Classical and quantum gravity models with N--connection on $T%
\mathbf{V}$ and/or $T^{\ast }\mathbf{V}$ \ (see details and references in
\protect\cite{vhfl,vstavr,av}):}

There is a recent interest in physics beyond the Standard Model which can
incorporate Lorentz symmetry violations, accelerating universe and dark
energy/matter effects in a Finsler setting which was discussed in above
subections, see also Refs. \cite%
{ika,gibbons,girelli,stavr1,stavr0,stavr2,sindoni,visser}. The Finsler
metrics are functions not only of the space-time coordinates but also of the
tangent vectors (momenta) at points of the curved manifolds. Certain ansatz for toy models 
were parametrized in diagonal form \cite{amelino,magueijo,kimberly,ellis}, or
with some examples of off--diagonal stochastic metrics \cite%
{mavromatos1,mavromatos2}. The fundamental issues of Finsler classical and
quantum gravity and cosmology related to N--connection and nonholonomic
structures were not studied in above mentioned references. In variables $%
(x^{i},p_{a}),$ and similar ones on higher order (co) tangent bundles, the problem of
constructing N--adapted classical and quantum gravity theories was studied
in \cite{vstavr,vsgg,av}. In such works, the velocity/momentum type
coordinates can be associated to (in general, higher order) spinor and/or
almost Kaehler variables which can can be used, for instance, for
deformation quantization of such theories.

Finally we note that geometrically all above mentioned model depending
constructions can be formalized using the concept of nonholonomic tangent
bundle/manifold which for nontrivial limits to standard GR should involve
nonholonomic deformations of Lorentz manifolds. A number of important
physical issues on fundamental property of cosmological models for such
locally anisotropic gravity theories (like viable cosmological models with
acceleration, diagonal and generic off--diagonal Finsler evolution etc) can
be studied following geometric and analytic methods of constructing exact
solutions, symmetries of such solutions and evolution scenarios.

\section{Accelerating Cosmology as Finsler Evolution}

\label{smfctb}

In this section, we show that the acceleration expansion of the present
matter--dominated universe may be generated along with the evolution of
Finsler space in velocity type dimensions. Two examples of exact
off--diagonal solutions associated with cosmological evolution scenarios
will be constructed. We prove that solitonic nonholonomic deformations
induced by velocity type variables modify scenarios of acceleration in real
Universe.

\subsection{Diagonal accelerating Finsler universes}

\label{ssacdiag} We consider a prime metric of a $(n+1+3)$--dimensional
spacetime, with $n=4$ and $m=1+3,$ with time like coordinate $y^{5}=t,$
{\small
\begin{eqnarray}
\ _{0}\mathbf{g}=\varepsilon _{1}dx^{1}\otimes dx^{1}+\frac{\ ^{h}a^{2}(t)}{%
1-\ ^{h}k(\ ^{h}r)^{2}}d\ ^{h}r\otimes d\ ^{h}r+\ ^{h}a^{2}(t)(\
^{h}r)^{2}d\ ^{h}\theta \otimes d\ ^{h}\theta  &&  \notag \\
+\ ^{h}a^{2}(t)\ (\ ^{h}r)^{2}\sin ^{2}\ ^{h}\theta \ d\ ^{h}\varphi \otimes
d\ ^{h}\varphi -dt\otimes dt+\frac{\ ^{v}a^{2}(t)}{1-\ ^{v}k(\ ^{v}r)^{2}}d\
^{v}r\otimes d\ ^{v}r &&  \notag \\
+\ ^{v}a^{2}(t)(\ ^{v}r)^{2}d\ ^{v}\theta \otimes d\ ^{v}\theta +\
^{v}a^{2}(t)\ (\ ^{v}r)^{2}\sin ^{2}\ ^{v}\theta \ d\ ^{v}\varphi \otimes d\
^{v}\varphi .\  &&  \label{diagfans}
\end{eqnarray}%
} This diagonal ansatz is considered for a "simple" cosmological model, with
zero N--connection coefficients, on a tangent bundle to a 4--d (pseudo)
Riemannian manifold, when $\widehat{\mathbf{D}}=\nabla .$ \footnote{%
The local coordinates are $u^{\alpha }=(x^{i},y^{a}),$ for $i,j,...=1,2,3,4$
and $a,b,...=5,6,7,8,$ (the time like coordinate $t$ is considered as the
first "fiber" coordinate) and the coefficients of $\ _{0}\mathbf{g}_{\alpha
\beta }=diag[\ _{0}g_{i},\ _{0}h_{a}]$ are, for spherical h--coordinates:\ $%
x^{1}=x^{1},x^{2}=\ ^{h}r,x^{3}=\ ^{h}\theta ,x^{4}=\ ^{h}\varphi ,$ and,
for v--coordinates, \ $y^{5}=t,y^{6}=\ ^{v}r,y^{7}=\ ^{v}\theta ,y^{8}=\
^{v}\varphi $; $\ _{0}g_{1}=\varepsilon _{1}=\pm 1,\ _{0}h_{5}=-1$,
\begin{eqnarray*}
\ _{0}g_{2} &=&\frac{\ ^{h}a^{2}(t)}{1-\ ^{h}k(\ ^{h}r)^{2}},\ _{0}g_{3}=\
^{h}a^{2}(t)(\ ^{h}r)^{2},\ _{0}g_{4}=\ ^{h}a^{2}(t)\ (\ ^{h}r)^{2}\sin
^{2}\ ^{h}\theta ; \\
\ _{0}h_{6} &=&\frac{\ ^{v}a^{2}(t)}{1-\ ^{v}k(\ ^{v}r)^{2}},\ _{0}h_{7}=\
^{v}a^{2}(t)(\ ^{v}r)^{2},\ \ _{0}h_{8}=\ ^{v}a^{2}(t)\ (\ ^{v}r)^{2}\sin
^{2}\ ^{v}\theta ,
\end{eqnarray*}%
} We study  a toy model on $T\mathbf{V}$ for a metric with
both the h-- and v--parts  of  FRW type, i.e.  spherical symmetries on h-
and  v--coordinates. The constructions can be involved in a class of solutions for  nonlinear deformed
gravity (in some sense, with "double" FRW "position" and "phase"
cosmology). To generate such solutions we have to chose  a source $\
\widehat{\mathbf{\Upsilon }}_{\beta \delta }$ on $T\mathbf{V}$ for the
generalized Einstein equations (\ref{ensteqcdc}), using N--adapted lifts
from $hT\mathbf{V}$ to $T\mathbf{V,}$ when a perfect fluid on the horizontal
part possess certain vertical anisotropies resulting in acceleration of
 observable h--subspace. We do not have experimental data
which would suggest  how a "perfect" v--fluid could move in such extra
dimensions (here we note that in the diagonal case, the N--connection is trivial). Nevertheless,   we
can approximate the dynamics of such locally anisotropic distributions of
matter to be with effective density function $\ ^{v}\underline{\rho }$ and
presure$\ ^{v}\underline{p}$ stated as "anisotropically polarized
cosmological constants in a 8--d bulk. It should be noted that the physical
interpretation of v--coordinates depend on the type locally anisotropic
theory we consider for our cosmological model (as we explained in section %
\ref{ssphint}). Geometrically, such constructions are for generic diagonal
solutions with "double" spherical symmetry of  (\ref{eq1})--(\ref{eq4})
which will be used as prime metrics for generating more realistic off--diagonal locally
anisotropic models, see  next section and discussions for  metrics (\ref{odyn4d}) and (\ref%
{solitsolut}).

\subsubsection{Diagonal cosmological equations for pseudo--Finsler metrics}

The metric (\ref{2forman}) describes two types (conventional horizontal and
vertical ones) evolutions with time variable $t$ of two universes
with respective constat curvatures $\ ^{h}k$ and $\ ^{v}k.$ To derive
cosmological solutions in a most simple form is convenient to consider in the 
 $h$--subspace a radial coordinate $\ 0<\
^{h}r<1$ taken for the light velocity $c=1.$ The coefficients $\ _{0}h_{a}$
define a usual FRW type metric in the $v$--subspace. The values $\ ^{h}a^{2}(t)$
and $\ ^{v}a^{2}(t)$ are respective $h$-- and $v$--scale factors. In such
models, the $h$--coordinates are dimensionless (we can introduce a standard
dimension, for instance, by multiplying on Planck length) and the $v$%
--coordinates are usual ones, with dimension of length.

Assuming that the matter content in this pseudo--Finsler spacetime is taken
to be a perfect fluid, we can write the Einstein equations (\ref{eqe1})--(%
\ref{eqe3}) as {\small
\begin{eqnarray}
4\frac{\ ^{v}a^{\bullet }}{\ ^{v}a}\frac{\ ^{h}a^{\bullet }}{\ ^{h}a}+2\left[
\left( \frac{\ ^{h}a^{\bullet }}{\ ^{h}a}\right) ^{2}+\frac{\ ^{h}k}{\left(
\ ^{h}a\right) ^{2}}\right] + \left[ \left( \frac{\ ^{v}a^{\bullet }}{\ ^{v}a%
}\right) ^{2}+\frac{\ ^{v}k}{\left( \ ^{v}a\right) ^{2}}\right] &=&\frac{8}{3%
}\pi \underline{G}\underline{\rho },  \label{dfrweq} \\
4\frac{\ ^{h}a^{\bullet \bullet }}{\ ^{h}a}+2\frac{\ ^{v}a^{\bullet \bullet }%
}{\ ^{v}a}+6\left[ \left( \frac{\ ^{h}a^{\bullet }}{\ ^{h}a}\right) ^{2}+%
\frac{\ ^{h}k}{\left( \ ^{h}a\right) ^{2}}\right] + \left[ \left( \frac{\
^{v}a^{\bullet }}{\ ^{v}a}\right) ^{2}+\frac{\ ^{v}k}{\left( \ ^{v}a\right)
^{2}}\right] &=&-8\pi \underline{G}\ ^{v}\underline{p},  \notag \\
\frac{\ ^{h}a^{\bullet \bullet }}{\ ^{h}a}+\frac{\ ^{v}a^{\bullet \bullet }}{%
\ ^{v}a}+2\left[ \left( \frac{\ ^{h}a^{\bullet }}{\ ^{h}a}\right) ^{2}+\frac{%
\ ^{h}k}{\left( \ ^{h}a\right) ^{2}}\right] + \left[ \left( \frac{\
^{v}a^{\bullet }}{\ ^{v}a}\right) ^{2}+\frac{\ ^{v}k}{\left( \ ^{v}a\right)
^{2}}\right] &=&-\frac{8}{3}\pi \underline{G}\ ^{h}\underline{p}.  \notag
\end{eqnarray}%
} In the above formulas, the right "dot" means derivative on time coordinate
$t $ and $\underline{G}$ and $\underline{\rho }$ are respectively the formal
gravitational constant and the energy density in the total (tangent) space.
The values $\ ^{h}\underline{p}$ and $\ ^{v}\underline{p}$ are,
correspondingly, the pressures in the $h$- and $v$--spaces. We assume simple
equations of matter states of type $\ ^{h}\underline{p}=\ \ ^{h}\omega \ ^{h}%
\underline{\rho }$ and $\ ^{v}\underline{p}=\ ^{v}\omega \ ^{v}\underline{%
\rho }$ for some constant state parameters $\ \ ^{h}\omega $ and $\
^{v}\omega $.\footnote{%
From a formal point of view, we can construct on tangent bundles perfect
fluid models with formal different h- and v--pressures, in N--adapted form
as we discussed in \cite{vsgg} (we omit in this work such details and send
the reader to a paper on "anisotropic inflationary model by S. Vacaru and D.
Gonta, in that collection of papers).} The conservation law $\nabla _{\alpha
}T^{\alpha \beta }=0$ for $T^{\alpha \beta }=diag[\ ^{h}\underline{\rho },\
^{h}\underline{p},...;\ ^{v}\underline{\rho },\ ^{v}\underline{p}.....]$
gives rise to $\underline{\rho }\eqsim \ ^{h}a^{-4(1+\ ^{h}\omega )}\times \
^{v}a^{-3(1+\ ^{v}\omega )}.$ For simplicity, we may assume $\ ^{h}k=0$ and
study the evolution of the scale factors $\ ^{h}a(t)$ and $\ ^{v}a(t)$ and
approximations in (\ref{dfrweq}).

\subsubsection{Diagonal scale evolution and velocity type dimensions}

In general, a Finsler gravity dynamics is with generic off--diagonal metrics
and generalized connections. Such nonlinear systems may result in
non--perturbative effects and instability even for small off--diagonal
metric terms.

\paragraph{Radiation--dominated diagonal Finsler universe:}

\ We define such an universe following conditions $\ ^{h}\underline{p}=\ 0$
and $\ ^{v}\underline{p}=\frac{1}{3}\ ^{v}\underline{\rho },$ when $\
^{h}a=const$ is accepted as a solution. For such configurations, the third
equation in (\ref{dfrweq}) is a consequence of the first two ones when $%
^{v}a(t)$ is a solution of equations
\begin{equation*}
\left( \frac{\ ^{v}a^{\bullet }}{\ ^{v}a}\right) ^{2}+\frac{\ ^{v}k}{\left(
\ ^{v}a\right) ^{2}} =\frac{8}{3}\pi \underline{G}\underline{\rho },\ \ 2%
\frac{\ ^{v}a^{\bullet \bullet }}{\ ^{v}a}+\left( \frac{\ ^{v}a^{\bullet }}{%
\ ^{v}a}\right) ^{2}+\frac{\ ^{v}k}{\left( \ ^{v}a\right) ^{2}} =-\frac{8}{3}%
\pi \underline{G}\underline{\rho }.
\end{equation*}%
The source $\frac{8}{3}\pi \underline{G}\underline{\rho }$ of such equations
is determined by generalized gravitational constant $\underline{G}$ and $%
\underline{\rho }$ matter density in total spacetime. By straightforward
computations, we can show that the constant $\ ^{h}a$--solution is stable
under small perturbations of scale factors $\ ^{h}a(t)$ and $\ ^{v}a(t).$
This means that we can retrieve the ordinary evolution of
radiation--dominated Finsler universe with a total spacetime model. Here we
note that for \ a matter--dominated Finsler configuration with $\ ^{h}%
\underline{p}=\ ^{v}\underline{p}=\ 0$ there is not a solution with $\
^{h}a=const$ unless $\underline{\rho }=0.$

\paragraph{Matter--dominated diagonal Finsler universe:}

\ There are solutions as in the standard FRW cosmology (in our case, for the
$v $--part) with $\ ^{h}a=const,$ when the matter in the ''velocity'' space
provides negative pressure $\ ^{h}\underline{p}=-\frac{1}{2}\underline{\rho }%
,$ when $\ ^{v}\underline{p}=0.$ Such conditions may be realistic if we
associate point like non--relativistic particles in $v$--space certain
extended objects (let say, strings) with additional velocity variables when
the pressure is provided in such a strange manner. The Friedman--Finsler
equations (\ref{dfrweq}) transform into $\left( \frac{\ ^{v}a^{\bullet }}{\
^{v}a}\right) ^{2}+\frac{\ ^{v}k}{\left( \ ^{v}a\right) ^{2}} =\frac{8}{3}%
\pi \underline{G}\underline{\rho }, \ \ 2\frac{\ ^{v}a^{\bullet \bullet }}{\
^{v}a}+\left( \frac{\ ^{v}a^{\bullet }}{\ ^{v}a}\right) ^{2}+\frac{\ ^{v}k}{%
\left( \ ^{v}a\right) ^{2}} =0,$ which allows us to find general solutions
for $\ ^{v}a(t).$

\paragraph{Different extension rates in $h$-- and $v$--subspaces:}

\ For simplicity, we can assume $\ ^{h}k=\ ^{v}k=0$ and that some constants $%
\ ^{h}\omega $ and $\ ^{v}\omega $ determine $\ ^{h}\underline{p}=\ \
^{h}\omega \ ^{h}\underline{\rho }$ and $\ ^{v}\underline{p}=\ ^{v}\omega \
^{v}\underline{\rho },$ i.e. the equations of states in a matter like
dominated Finlser universe. The difference between the $v$- and $h$%
--expansion rates is expressed $\ _{h}^{v}\beta (t) :=(1-3\ ^{v}\omega +2\
^{h}\omega )\frac{\ ^{v}a^{\bullet }}{\ ^{v}a}-\left[ 1+3\ ^{v}\omega -4\
^{h}\omega \right] \frac{\ ^{h}a^{\bullet }}{\ ^{h}a} \eqsim \frac{1}{(\
^{v}a)^{3}\ (\ ^{h}a)^{4}}\eqsim \frac{1}{Vol_{3+4}},$ where $Vol_{3+4}$ is
the volume of a $(3+4)$--dimensional space like total pseudo--Finsler
subspace if signature $\varepsilon _{1}=1$ in (\ref{dfrweq}). It follows
from this formula that the difference $\ _{h}^{v}\beta (t)$ decreases
(grows) as the volume $Vol_{3+4}$ grows (decreases). Here we note that $\
(^{h}a)^{4}$ has a limit corresponding to the maximal velocity of light.

Similarly, it is possible to consider the difference between $v$- and $h$%
--expansions $\ _{h}^{v}\beta (t)$ for a radiation--dominated Finsler
universe with $\ ^{h}\underline{p}=\frac{1}{3}\underline{\rho }$ and $\ ^{v}%
\underline{p}=0,$ $\ _{h}^{v}\beta (t):=2\frac{\ ^{h}a^{\bullet }}{\ ^{h}a}%
\eqsim \frac{1}{(\ ^{v}a)^{3}\ (^{h}a)^{4}}\eqsim \frac{1}{Vol_{3+4}}.$ We
conclude that if $Vol_{3+4}$ is growing, the expansion rate of the $h$%
--spaces drops to zero. So, the constant $\ ^{h}a$ solution is stable for
the radiation dominated Finsler universe.

It is possible to consider a more general matter dominated Finsler universe
with $\ \ ^{h}\omega =\ ^{v}\omega \neq \frac{1}{3}$ when $\ _{h}^{v}\beta
(t)\eqsim \frac{\ ^{v}a^{\bullet }}{\ ^{v}a}-\frac{\ ^{h}a^{\bullet }}{\
^{h}a}\eqsim \frac{1}{(\ ^{v}a)^{3}\ (^{h}a)^{4}}\eqsim \frac{1}{Vol_{3+4}}.$

If the total volume $Vol_{3+4}$ is growing, the expansion rates of the $h$-
and $v$--spaces tend to approach each other, i.e. the limited $h$--volume,
because of finite speed of light, limits the three--space in the $v$--part.
If for an inverse decreasing of $Vol_{3+4},$ with one expanding and another
collapsing subspaces, then $\ |\ ^{v}a^{\bullet }/\ ^{v}a|,$ \ or $\ |\
^{h}a^{\bullet }/\ ^{h}a|,$ \ becomes large and larger. This results in an
accelerating expansion. For collapsing $h$--space with velocity types
coordinates we can induce an accelerating expansion of ''our'' inverse
modelled by this pseudo--Finsler model as the $v$--subspace. We analyze
below more details on such models of Finsler--acceleration.

\subsubsection{Accelerating diagonal expansion with Finsler evolution}

We explore analytically the possibility to generating Finsler type
accelerating expansions via evolution of $h$--space with velocity
coordinates. For simplicity, we consider $\ ^{h}k=\ ^{v}k=0$ and trivial
equations of states with $\ ^{h}\underline{p}=\ ^{v}\underline{p}=0.$%
\footnote{%
Models with nonzero$\ ^{h}k$ and/or$\ ^{v}k$ offer a number of interesting
possibilities. In the next subsections, we shall investigate examples with
nontrivial N--connection and Riemannian and scalar curvatures.} For such
conditions, the last two equations in (\ref{dfrweq}) became {\small \ $\
^{v}H^{\bullet }+\frac{5}{2}(\ ^{v}H) ^{2}+2\ ^{v}H\ \ ^{h}H-(\ ^{h}H) ^{2}
=0,\ \ ^{h}H^{\bullet }-\frac{1}{2}(\ ^{v}H) ^{2}+\ ^{v}H\ \ ^{h}H+3 (\
^{h}H) ^{2} =0,$ } where the respective effective Hubble $h$-- and $v$%
--''constants'' are $\ ^{h}H:=\ ^{h}a^{\bullet }/\ ^{h}a$ and $\ ^{v}H:=\
^{v}a^{\bullet }/\ ^{v}a.$ These equations impose the corresponding
conditions for accelerating {\small ($\ ^{v}a^{\bullet \bullet }/\ ^{v}a>0$)}%
, or decelerating {\small ($\ ^{v}a^{\bullet \bullet }/\ ^{v}a<0$)} of
''our'' three dimensional $v$--subspace, acceleration:\ {\small $\ ^{h}H >
\left( 1+\sqrt{5/2}\right) \ ^{h}H:=\ ^{+}H\ \ ^{h}H,$ or $\ ^{h}H > \left(
1-\sqrt{5/2}\right) \ ^{h}H:=\ ^{-}H\ \ ^{h}H$}; \ deceleration: {\small \ $%
\ ^{-}H\ \ ^{v}H < \ ^{h}H<\ ^{+}H\ \ ^{v}H$}. To investigate the
correlation between $h$-- and $v$--subspaces is useful to introduce the
fraction--function $\gamma (t):=\ \ ^{h}H/\ \ ^{v}H$ and see the behavior of
$d\gamma /dt$ for different values of $\gamma $ and some ''critical'' values
of this function, which for our dimensions $n=4$ and $m=1+3$ are defined:\
attracting:\ $\ _{att}H:=-1+1/\sqrt{2}$; repelling:\ $\ _{rep}H:=-1-1/\sqrt{2%
}$. It is always satisfied the condition $\ _{att}H<\ ^{-}H<\ _{rep}H<0<1<\
^{+}H,$ i. e. there are two ''attractors'' determined by $\gamma =\ _{att}H$
and $\gamma =1$ and one ''repeler'' for $\gamma =\ _{rep}H.$ One follows the
conditions $\gamma ^{\bullet } >0,$ for $\gamma < \ _{att}H,\ \
_{rep}H<\gamma <1; \gamma ^{\bullet } < 0,$ for $\ _{att}H < \gamma < \
_{rep}H,\gamma > 1$.

For Finsler universes, there are four kinds of evolution processes depending
of a initial value $\gamma =\ ^{\circ }\gamma :$%
\begin{eqnarray}
\mbox{ acceleration and, then, deceleration,  }\ ^{\circ }\gamma &>&\ ^{+}H;
\label{evolutcond} \\
\mbox{ always deceleration, }\ _{rep}H &<&\ ^{\circ }\gamma <\ ^{+}H;  \notag
\\
\mbox{ deceleration and, then acceleration, }\ ^{-}H &<&\ ^{\circ }\gamma <\
_{rep}H;  \notag \\
\mbox{ always acceleration, }\ ^{\circ }\gamma &<&\ ^{-}H.  \notag
\end{eqnarray}%
A realistic for our universe is the third condition above, when $\ ^{-}H<\
^{\circ }\gamma <\ _{rep}H.$ Such a scenario states that initially the
Finsler universe is in the region $\left( \ ^{-}H,\ _{rep}H\right) $ when
the $h$--space collapses and our three dimension space in the $v$--part
decelerates. As $\ \gamma $ passes $\ ^{-}H$ in the collapsing process of
''velocity'' $h$--coordinates, our ''real'' 3--d space begins to accelerate.

\subsection{Off--diagonal anisotropic Finsler acceleration}

\label{ssofdfa} More realistic Finsler type cosmological models can be
elaborated for generic off--diagonal metrics and with nontrivial
N--connection.

\subsubsection{Examples of off--diagonal cosmological solutions}

We construct in explicit form two classes of such solutions defining certain
models of four dimensional, 4-d, and 8-d Finsler spacetimes.

\paragraph{A pseudo--Finsler 4-d off--diagonal toy cosmology:}

\ Let us consider
\begin{eqnarray}
\mathbf{\check{g}} &=&\ ^{h}a^{2}(t)d\ ^{h}r\otimes d\ ^{h}r-dt\otimes dt+\
^{h}a^{2}(t)(\ ^{h}r)^{2}d\ ^{h}\theta \otimes d\ ^{h}\theta  \notag \\
&&+\ ^{h}a^{2}(t)\ (\ ^{h}r)^{2}\sin ^{2}\ ^{h}\theta \ d\ ^{h}\varphi
\otimes d\ ^{h}\varphi ,  \label{dyn4d}
\end{eqnarray}%
which is contained as a particular case of 8-d ansatz (\ref{diagfans}), when
$\varepsilon _{1}=0,\ ^{v}a=0$ an, for simplicity, $\ ^{h}k=\ ^{v}k=0.$ We
use this metric for a prime cosmological model in variables $u^{\widehat{%
\alpha }}=(\ ^{h}r,t,\ ^{h}\theta ,\ ^{h}\varphi ),$ with $x^{\widehat{i}%
}=(\ ^{h}r,t)$ and $y^{\widehat{a}}=(\ ^{h}\theta ,\ ^{h}\varphi ),$ for $%
\widehat{i},\widehat{j},...=2,3$ and $\widehat{a},\widehat{b},...=4,5$ (such
a model describes evolution in time $t$ of a $h$-- subspace for certain
conditions, and sources, analyzed in subsection \ref{ssacdiag}). An
off--diagonal anisotropic dynamics in the space of ''velocities''\ can be
modelled by nonholonomic deformations with $\eta $--polarizations $\eta _{%
\widehat{i}}=\eta _{\widehat{i}}(x^{\widehat{k}})$ and $\eta _{\widehat{a}%
}=\eta _{\widehat{a}}(x^{\widehat{k}},\ ^{h}\theta )$ and N--connection
coefficients $N_{\widehat{i}}^{\widehat{4}}=w_{\widehat{i}}(x^{\widehat{k}%
},\ ^{h}\theta )$ and $N_{\widehat{i}}^{\widehat{5}}=n_{\widehat{i}}(x^{%
\widehat{k}},\ ^{h}\theta ).$ For the ''prime'' metric {\small $\check{g}_{%
\widehat{2}}=\ ^{h}a^{2}(t), \check{g}_{\widehat{3}}=-1,\check{h}_{\widehat{4%
}}=\ ^{h}a^{2}(t)(\ ^{h}r)^{2},\check{h}_{\widehat{5}}=\ ^{h}a^{2}(t)(\
^{h}r)^{2}\sin ^{2}\ ^{h}\theta ,$ we define$\ \mathbf{\check{g}}=\left[
\check{g}_{\widehat{i}},\check{h}_{\widehat{a}},\check{N}_{\widehat{j}}^{%
\widehat{a}}\right] \rightarrow \mathbf{g=}\left[ g_{\widehat{i}}=\eta _{%
\widehat{i}}\check{g}_{\widehat{i}},h_{\widehat{a}}=\eta _{\widehat{a}}%
\check{h}_{\widehat{a}},N_{\widehat{i}}^{\widehat{a}}\right] ,$} to a metric
\begin{equation}
\ ^{4d}\mathbf{g}=g_{\widehat{i}}d\ x^{\widehat{i}}\otimes dx^{\widehat{i}%
}+h_{\widehat{a}}(dy^{\widehat{a}}+N_{\widehat{i}}^{\widehat{a}}dx^{\widehat{%
i}})\otimes (dy^{\widehat{a}}+N_{\widehat{i}}^{\widehat{a}}dx^{\widehat{i}}),
\label{odyn4d}
\end{equation}%
constrained to be a cosmological solution of equations (\ref{ep1a})--(\ref%
{ep4a}) for $\widehat{\mathbf{D}}.$ For simplicity, we consider a source
with constant coefficients $\Upsilon _{\ \quad \widehat{\beta }}^{\ \widehat{%
\alpha }}=diag[\Upsilon _{\ \widehat{\gamma }};$ $\Upsilon _{2}=\Upsilon
_{3}=const;\Upsilon _{4}=\Upsilon _{5}=const]$ transforming for a
''diagonal'' limit into $T^{\alpha \beta }=diag[\ ^{h}\underline{\rho },\
^{h}\underline{p},...;\ ^{v}\underline{\rho },\ ^{v}\underline{p}.....]$
used for generating a metric (\ref{dfrweq}).

The coefficients of such a new solution (\ref{odyn4d}) are of type (\ref%
{sol1}) generated from (\ref{dyn4d}) by polarization functions and
N--connection coefficients\footnote{%
to simplify formulas, we chose corresponding parametrizations for
generating/ integration functions} : {\small
\begin{eqnarray}
\eta _{2} &=&e^{\psi (\ ^{h}r,t)}\ ^{h}a^{-2}(t),\ \eta _{3}=e^{\psi (\
^{h}r,t)}\mbox{\ for }\psi ^{\bullet \bullet }-\psi ^{\prime \prime
}=\Upsilon _{2};  \notag \\
\eta _{4} &=&\ [f^{\ast }(\ ^{h}r,t,\ ^{h}\theta )]^{2}|\varsigma (\
^{h}r,t,\ ^{h}\theta )|,  \label{odsol1} \\
&&\mbox{for }\varsigma =\ 1-\frac{\Upsilon _{4}}{8}\int d\ ^{h}\theta \
f^{\ast }(\ ^{h}r,t,\ ^{h}\theta )\ [f(\ ^{h}r,t,\ ^{h}\theta )-\ ^{0}f(\
^{h}r,t)],  \notag \\
\eta _{5} &=&[f(\ ^{h}r,t,\ ^{h}\theta )-\ ^{0}f(\ ^{h}r,t)]^{2};  \notag \\
w_{\widehat{j}} &=&\ _{0}w_{\widehat{j}}(\ ^{h}r,t)\exp \{ -\int_{0}^{\
^{h}\theta }\left[ \frac{2\eta _{4}A^{\ast }}{\eta _{4}^{\ast }}\right]
_{v\rightarrow v_{1}}dv_{1}\} \int_{0}^{^{h}\theta } dv_{1}\left[ \frac{\eta
_{4}B_{j}}{\eta _{4}^{\ast }}\right] _{v\rightarrow v_{1}}  \notag \\
&& \exp \{ -\int_{_{0}}^{v_{1}} [\frac{2\eta _{4}A^{\ast }}{\eta _{4}^{\ast }%
}] _{v\rightarrow v_{1}}dv_{1} \},\ n_{\widehat{i}} =\ _{0}n_{\widehat{i}}(\
^{h}r,t)+\int d\ ^{h}\theta \ \eta _{4}\ ^{h}a^{2}(t)(\ ^{h}r)K_{\widehat{i}%
},  \notag
\end{eqnarray}%
} where the coefficients of type (\ref{aux}) are computed for polarization
functions,{\small
\begin{eqnarray}
A &=&\left( \frac{\eta _{4}^{\ast }}{2\eta _{4}}+\frac{\eta _{5}^{\ast }}{%
2\eta _{5}}\right) ,\ B_{\widehat{k}}=\frac{\eta _{5}^{\ast }}{2\eta _{5}}%
\left( \frac{\partial _{\widehat{k}}g_{\widehat{2}}}{2g_{\widehat{2}}}-\frac{%
\partial _{\widehat{k}}g_{\widehat{3}}}{2g_{\widehat{3}}}\right) -\partial _{%
\widehat{k}}A,  \notag \\
K_{2} &=&-\frac{1}{2}\left( \frac{g_{2}^{\prime }}{g_{3}h_{5}}+\frac{%
g_{3}^{\bullet }}{g_{3}h_{5}}\right) ,\ K_{3}=\frac{1}{2}\left( \frac{%
g_{3}^{\bullet }}{g_{2}h_{4}}-\frac{g_{3}^{\prime }}{g_{3}h_{5}}\right) .
\label{nondiag}
\end{eqnarray}
}In formulas (\ref{odsol1}) and (\ref{nondiag}), the partial derivatives are
written in brief in the form $\eta _{4}^{\ast }=\partial \eta _{4}/\partial
\ ^{h}\theta ,\ $\ $g_{3}^{\bullet }=\partial g_{3}/\partial \ ^{h}r,$ $%
g_{3}^{\prime }=\partial g_{3}/\partial t$ and $\ ^{0}f(\ ^{h}r,t),\ _{0}w_{%
\widehat{j}}(\ ^{h}r,t),$ $\ _{0}n_{\widehat{i}}(\ ^{h}r,t)$ are integration
functions to be determined by fixing some boundary/initial conditions in the
space of ''velocities''.

Putting together the above coefficients, we find the 4-d metric {\small
\begin{eqnarray*}
&&\mathbf{g} = e^{\psi(\ ^{h}r,t)} (d\ ^{h}r\otimes d\ ^{h}r-dt\otimes dt)+
[f^{\ast}(\ ^{h}r,t,\ ^{h} \theta )]^{2}|\varsigma (\ ^{h}r,t,\ ^{h}\theta
)|\ ^{h}a^{2}(t) (\ ^{h}r)^{2} \\
&& \delta \ ^{h}\theta \otimes \delta \ ^{h}\theta +[f(\ ^{h}r,t,\
^{h}\theta )-\ ^{0}f(\ ^{h}r,t)]^{2}\ ^{h}a^{2}(t)\ (\ ^{h}r)^{2}\sin ^{2}\
^{h}\theta \delta \ ^{h}\varphi \otimes \delta \ ^{h}\varphi , \\
&&\delta \ ^{h}\theta =d\ ^{h}\theta +w_{\widehat{2}}(\ ^{h}r,t,\ ^{h}\theta
)d\ ^{h}r+w_{\widehat{3}}(\ ^{h}r,t,\ ^{h}\theta )d\ ^{h}r, \\
&&\delta \ ^{h}\varphi =d\ ^{h}\varphi +n_{\widehat{2}}(\ ^{h}r,t,\
^{h}\theta )d\ ^{h}r+n_{\widehat{3}}(\ ^{h}r,t,\ ^{h}\theta )d\ ^{h}r,
\end{eqnarray*}%
} with the coefficients defined by data (\ref{odsol1}). Such a metric
defined an off--diagonal Finsler inhomogeneous model in the $h$--subspace.
In our ''real'' Universe it may contribute via a nontrivial $e^{\psi (\
^{h}r,t)}$ before time like $dt$; such a solution should be imbedded into a
8--d Finsler spacetime.

\paragraph{A class of inhomogeneous off--diagonal 8--d Finsler cosmologies:}

\ Following the geometric method of constructing exact solutions in extra
dimensional spacetime \cite{vgensol}, we can generalize the metric (\ref%
{odyn4d}) with coefficients (\ref{odsol1}) to generate cosmological
solutions for a total 8-d Finsler spacetime. We consider a source (\ref{s8d}%
) with constant coefficients modeling on $h$-- and $v$--subspaces perfect
fluid matter/radiation states. The 8-d ansatz is {\small
\begin{eqnarray}
\ ^{8d}\mathbf{g} &=&\varepsilon _{1}dx^{1}\otimes dx^{1}+\eta _{\widehat{i}}%
\check{g}_{\widehat{i}}d\ x^{\widehat{i}}\otimes dx^{\widehat{i}}+\eta _{4}%
\check{h}_{4}(dy^{4}+w_{\widehat{i}}dx^{\widehat{i}})\otimes (dy^{4}+w_{%
\widehat{i}}dx^{\widehat{i}})  \notag \\
&&+h_{5}(dy^{5}+w_{\alpha }(u^{\alpha },\ ^{1}v)du^{\alpha })\otimes
(dy^{5}+w_{\alpha }(u^{\alpha },\ ^{1}v)du^{\alpha })  \notag \\
&&+h_{6}(dy^{6}+n_{\alpha }(u^{\alpha },\ ^{1}v)du^{\alpha })\otimes
(dy^{6}+n_{\alpha }(u^{\alpha },\ ^{1}v)du^{\alpha })  \label{8dods} \\
&&+h_{7}(dy^{7}+w_{\ ^{1}\alpha }(u^{\ ^{1}\alpha },\ ^{2}v)du^{^{1}\alpha
})\otimes (dy^{7}+w_{\ ^{1}\alpha }(u^{\ ^{1}\alpha },\ ^{2}v)du^{^{1}\alpha
})  \notag \\
&&+h_{8}(dy^{8}+n_{\ ^{1}\alpha }(u^{\ ^{1}\alpha },\ ^{2}v)du^{^{1}\alpha
})\otimes (dy^{8}+n_{\ ^{1}\alpha }(u^{\ ^{1}\alpha },\ ^{2}v)du^{^{1}\alpha
}),  \notag
\end{eqnarray}%
} where coordinates and respective indices are parameterized $u^{\alpha }
=(x^{1},x^{\widehat{i}}=(\ ^{h}r,t),x^{4}=y^{4}=\ ^{h}\theta );\ u^{\
^{1}\alpha } = (u^{\alpha },y^{\ ^{1}a}=(y^{\ 5}=\ ^{1}v=\ ^{v}r,y^{\ 6}=\
^{h}\varphi ));\ u^{\ ^{2}\alpha } = (u^{\ ^{2}\alpha },y^{\ ^{2}a}=(y^{\
7}=\ ^{2}v=\ ^{v}\theta ,y^{\ 8}=\ ^{v}\varphi ));$ and $h_{6}=\eta _{6}%
\check{h}_{6}+\underline{h}_{6}$ (for $\underline{h}_{6}$ known for given $%
h_{6}$ and $\eta _{6}\check{h}_{6}$), $n_{\alpha }(u^{\alpha },\ ^{1}v)=(n_{%
\widehat{i}}(x^{\widehat{k}},\ ^{h}\theta );n_{\alpha }(u^{\alpha },\
^{1}v), $ if $\ \alpha >3)$ when the values $\eta _{\widehat{i}}\check{g}_{%
\widehat{i}},\eta _{4}\check{h}_{4},\eta _{6}\check{h}_{6}$ being former $%
\eta _{5}\check{h}_{5}$ in (\ref{odsol1})$,$ $w_{\widehat{i}}$ and $n_{%
\widehat{i}}$ are given by coefficients of metrics (\ref{odyn4d}) and (\ref%
{dyn4d}).\footnote{%
In "standard" cosmological models based on GR, it is also possible to define
inhomogeneous and anisotropic configurations. For such cosmological
spacetimes, the metrics do not depend explicitly on "fiber/velocity" type
variables. Finsler anisotropic/inhomogeneous constructions are defined by
more complex diagonal and/or off--diagonal metrics on spacetimes with
tangent bundle structure.}

From the class of general solutions, we can extract a subclass of 3-d
solitonic configurations $\xi =\xi (t,\ ^{h}\theta ,\ ^{v}r)$ from the $h$%
--subspace, depending on time and velocity type coordinates, inducing small
perturbations in the $v$--subspace. Such anisotropic on velocities 8--d
metrics are written {\small
\begin{eqnarray}
\ ^{sol}\mathbf{g} &=&\varepsilon _{1}dx^{1}\otimes dx^{1}+e^{\psi (\
^{h}r,t)}\ ^{h}a^{2}(t)\ d\ ^{h}r\otimes d\ ^{h}r  \notag \\
&&+\eta _{4}(\ ^{h}r,t,\ ^{h}\theta )\ ^{h}a^{2}(t)(\ ^{h}r)^{2}\ \delta \
^{h}\theta (\ ^{h}r,t,\ ^{h}\theta )\otimes \delta \ ^{h}\theta (\ ^{h}r,t,\
^{h}\theta )  \notag \\
&&+\widetilde{\eta }_{6}[\xi ]\eta _{5}(\ ^{h}r,t,\ ^{h}\theta )\
^{h}a^{2}(t)(\ ^{h}r)^{2}\sin ^{2}\ ^{h}\theta \ \delta \ ^{h}\varphi
\lbrack \xi ]\otimes \delta \ ^{h}\varphi \lbrack \xi ]  \notag \\
&&-e^{\psi (\ ^{h}r,t)}dt\otimes dt+\left( 1+\varepsilon \varpi _{5}[\xi
]\right) \ ^{v}a^{2}(t)\ \delta \ ^{v}r[\xi ]\otimes \delta \ ^{v}r[\xi ]
\label{solitsolut} \\
&&+\ ^{v}a^{2}(t)(\ ^{v}r)^{2}\ d\ ^{v}\theta \otimes d\ ^{v}\theta +\
^{v}a^{2}(t)(\ ^{v}r)^{2}\sin ^{2}(\ ^{v}\theta )\ d\ ^{v}\varphi \otimes d\
^{v}\varphi ,  \notag \\
\mbox{where}&& \delta \ ^{h}\theta (\ ^{h}r,t,\ ^{h}\theta ) =d\ ^{h}\theta
+w_{2}(\ ^{h}r,t,\ ^{h}\theta )d\ ^{h}r+w_{3}(\ ^{h}r,t,\ ^{h}\theta )dt,
\notag \\
\delta \ ^{v}r[\xi ] &=&d\ ^{v}r+\varepsilon \widetilde{w}_{3}[\xi
]dt+\varepsilon \widetilde{w}_{4}[\xi ]d\ ^{h}\theta ,\ \delta \ ^{h}\varphi
\lbrack \xi ] =d\ \ ^{h}\varphi +\varepsilon \widetilde{n}_{3}[\xi
]dt+\varepsilon \widetilde{n}_{4}[\xi ]d\ ^{h}\theta ,  \notag
\end{eqnarray}%
} for values $e^{\psi },$ $\eta _{4},$ $w_{2},w_{3}$ and $\eta _{5}$
determined by formulas (\ref{odsol1}) and the coefficients depending
functionally on $[\xi ],$ for a 3--d solitonic function $\xi =\xi (t,\
^{h}\theta ,\ ^{v}r)$ (for instance, being a solitonic solution as we
computed the end of Appendix \ref{asect2}). Such 3-d solitons were
considered in our works on propagation of black holes in extra dimensional
spacetimes and on local anisotropic black holes in noncommutative gravity
\cite{vsingl,vncg,vncbh}. Solitonic configurations can be stable and
propagate from the space of ''velocities'' into ''real'' universe for
various models of Finsler cosmology.

For $\varepsilon \rightarrow 0,$ the solutions (\ref{solitsolut}) transform
into the 8-d metric (\ref{diagfans}) with possible nonholonomic
generalizations containing solutions of type (\ref{odyn4d}), (\ref{odsol1}).

\subsubsection{Solitonic N--connections and anisotropic acceleration}

The pseudo--Finsler cosmological model described by (\ref{solitsolut}) is
generically off--diagonal. The solitonic deformation $\xi $ contributes both
to diagonal and off--diagonal terms of metric. We can fix a nonholonomic(co)
frame of reference, $e^{\alpha }=(dx^{1},d\ ^{h}r,\delta \ ^{h}\theta
,\delta \ ^{h}\varphi ,dt,\delta \ ^{v}r,d\ ^{v}\theta ,d\ ^{v}\varphi ),$
for an observer in a point $\left( \ ^{v}r_{0},\ ^{v}\theta _{0},\
^{v}\varphi _{0}\right) ,$ for simplicity, considering that the ''velocity''
space is with $\ ^{h}r=0,$ $\ ^{h}\theta $ with one anisotropic velocity $\
^{h}\varphi .$ There are two effective scaling parameters $\ ^{h}\widetilde{a%
}(\tau )=\left( 1+\varepsilon \chi (\tau )\right) \ ^{h}a(\tau );\ ^{v}%
\widetilde{a}(\tau )=\left( 1+\varepsilon \varpi _{5}[\xi (\tau )]\right) \
^{v}a(\tau ),$ when we approximate $e^{\psi (\ ^{h}r_{0},\tau
)}=1+\varepsilon \chi (\tau ) $ with the solitonic function $\xi (\tau )$
taken for a redefined time like variable $\tau (t)$ when $d\tau =e^{\psi (\
0,t)}dt.$

Let us introduce {\small \ $\ ^{h}\widetilde{H}^{\star } :=\ ^{h}\widetilde{a%
}^{\star }/\ ^{h}\widetilde{a}=\ ^{h}\underline{H}+\varepsilon \chi ^{\star
},$ \ for $\ ^{h}\underline{H}=\ ^{h}a^{\star }/\ ^{h}a,$ $\ ^{v}\widetilde{H%
}^{\star } :=\ ^{v}\widetilde{a}^{\star }/\ ^{v}\widetilde{a}=\ ^{v}%
\underline{H}+\varepsilon \varpi _{5}^{\star },$ for $\ ^{v}\underline{H}=\
^{h}a^{\star }/\ ^{h}a,$ where $\chi ^{\star }=\partial \chi /\partial \tau
. $} The conditions for acceleration, in our case modified by a solitonic
function both for off--diagonal and diagonal terms of metric, are redefined
in the form {\small $\ ^{h}H\rightarrow \ ^{h}\widetilde{H},\
^{v}H\rightarrow \ ^{v}\widetilde{H}$ and $\partial /\partial t\rightarrow
\partial \tau .$} This introduces additional, solitonic type, correlations
between $h$-- and $v$--subspaces via an additional nonholonomic deformation
of fraction--function, {\small \ $\widetilde{\gamma }(\tau ):=\ \ ^{h}%
\widetilde{H}/\ \ ^{v}\widetilde{H}\approx \underline{\gamma }(\tau )\
+\varepsilon \left( \chi -\varpi _{5}\right) ^{\star }$.} Conclusions about
Finsler--solitonic off--diagonal acceleration, or deceleration, should be
drawn from behavior of functions $\widetilde{\gamma }^{\star }(\tau )$ and $%
\widetilde{\gamma }(\tau ).$ Respectively, the solitonic versions of
equations (\ref{evolutcond}) are {\small \ $\widetilde{\gamma }^{\star } > 0,%
\mbox{  for \ }\widetilde{\gamma }<_{att}H,\ \ _{rep}H<\widetilde{\gamma }%
<1; \ \widetilde{\gamma }^{\star } <0,\mbox{ for \ }\ _{att}H<\widetilde{%
\gamma }<\ _{rep}H,\ \widetilde{\gamma }>1,$ and (the four types of
evolution processes depend on a initial value $\widetilde{\gamma }=\ ^{\circ
}\widetilde{\gamma }$)%
\begin{eqnarray*}
\mbox{ acceleration and, then, deceleration,  }\ \ ^{\circ }\widetilde{%
\gamma } &>&\ ^{+}H; \\
\mbox{ always deceleration, }\ _{rep}H &<&\ \ ^{\circ }\widetilde{\gamma }<\
^{+}H; \\
\mbox{ deceleration and, then acceleration, }\ ^{-}H &<&\ ^{\circ }%
\widetilde{\gamma }<\ _{rep}H; \\
\mbox{ always acceleration, }\ ^{\circ }\widetilde{\gamma } &<&\ ^{-}H.
\end{eqnarray*}%
} A solitonically modified directly observable universe is $\ ^{-}H<\
^{\circ }\widetilde{\gamma }<\ _{rep}H.$ Such a condition is very sensitive
with respect to possible (off--) diagonal perturbations from the space of
velocities. This follows from the facts that the conditions of acceleration
for $\underline{\gamma }$ and $\widetilde{\gamma }=\underline{\gamma }\
+\varepsilon \left( \chi -\varpi _{5}\right) ^{\star }$ are, in general,
different. Small modifications proportional to $\varepsilon \left( \chi
-\varpi _{5}\right) ^{\star }$ may transfer, for instance, an accelerating
configuration into decelerating, and inversely.

In this section we have investigated the scenario of producing the
accelerating expansion of the present universe via evolving small velocity
type diagonal and off--diagonal nonholonomic deformations of Finsler
metrics. For a radiaton--dominated cosmological model, such as the model of
our early universe, we obtain stable configurations with static velocity
type coordinates. In this case, the existence of Finsler type velocity
coordinates may have no significant influence on 3--d observable space. Here
we note that on the contrary, diagonal solutions with static extra
dimensions does not exist for the present matter--dominated cosmologies.

There are four classes of evolution for the matter--dominated cases, as we
derived from our quantitative analysis of both types diagonal and
off--diagonal solutions. Cosmological models that decelerate first and than
accelerate are included into the schemes. Therefore the accelerating Finsler
expansion of the present universe may be described in our locally
anisotropic scenario. We note that small solitonic type deformations from
the velocity type subspace may modify substantially the character of
acceleration of universe and conditions of stability or instability.

Let us discuss several important properties of locally anisotropic (Finsler
type) cosmological solutions constructed in this section. There are
substantial differences if we compare our results with those obtained for
diagonal configurations in \cite{amelino,magueijo,kimberly}. Our
cosmological spacetimes are, in general, with nontrivial N--connection
structure, i.e. define nontrivial Finsler cosmologies which results in  anisotropically accelerating universes. For off--diagonal Finsler
cosmological solutions, we can can define commutative and/or noncommutative
black hole/ellipsoid/wormhole (or extra dimensional off--diagonal
cosmological) configurations as in \cite{vncg,vsingl,vcosm1}. Such exact
solutions can be constructed in explicit form for various models of
trapping/warping of Finsler branes and anisotropies of
Ho\v rava--Lifshitz--Finsler type \cite{vbranesf,vncbh,vhfl}.

Finally, we note that  off--diagonal metrics of type (\ref{odyn4d}) and (\ref%
{solitsolut}) posses certain similarities with metrics  (3.10) and (4.18)
constructed in Ref. \cite{mavromatos2} for a different type of  quantum
gravity phenomenology and accelerating cosmology derived for stochastic D--particle foam etc. Even the
string--brane stochastic models are constructed from another fundamental
mathematic and physical principles than those considered in this work, for
metric compatible Finsler gravity models, they can be unified in low energy
limits using generic off--diagonal stochastic solutions of generalized
Einstein equations as in Ref. \cite{vstoch}. In both cases of stochasic
D--particle foam and exact solutions with stochastic generating function the
cosmological  aspects of theories are related to anisotoropic modifications
of the Boltzmann equation (to get results similar to \cite%
{mavromatos1,mavromatos2} we should consider solutions from \cite{vstoch}
when the v--coordinates are considered for  distorted metric with momentum
transfer). Solutions of the Boltzmann equations and corresponding generating
functions or foam--modified thermal dark matter relic abundances can be
connected to  Standard Cosmology in certain trivial N--connection limits,
and absence of the foam.

\section{Concluding remarks}

\label{concl}

To avoid repetition, in this concluding section we do not attempt to
summarize all of the issues and application we discussed. This is because
as, in most cases, the preliminary insight gained and perspectives can not
be summarized in a sentence or two and this would not be very helpful to
readers. However, we have encountered two key issues: 1) The
Einstein--Finsler gravity can be formulated following the same principles as
general relativity but on certain nonholonomic bundle/manifold spaces and
corresponding generalized Finsler connections (which are also uniquely
defined by the coefficients of metric tensor in a metric compatible form).
2) Analyzing possible implications of quantum gravity and related Lorentz
violations in Early Universe and present day cosmology, we derived very easy
that the dynamics (in general, with nonholonomic constraints) in the space
of velocities contributes substantially to stability and
acceleration/deceleration stadia of cosmological models.

The main purpose of this work was to show how the main postulates for the
general relativity theory can be extended on nonholonomic tangent bundles/
manifolds. It was provided a self--consistent scheme for formulating Finsler
gravity models and fundamental physical equations in a form most closed to
standard particle physics. Applying the anholonomic deformation method, it
was possible to construct new classes of exact cosmological solutions with
generic off--diagonal metrics. We also analyzed  scenarios for Finsler
acceleration of Universe.

One of the most important cosmological problem  which should be solved in
Finsler type gravity theories is that how we could avoid possible
overclsoure of the Universe. This non--trivial issue  was not
studied in this paper. We suppose that realistic cosmological  scenarios can be elaborated in
a self--consistent theoretical form and in correspondence with observational
data for a certain Finsler brane cosmological models with trapping/workping
as we discussed in \cite{vbranesf,vncbh,vhfl}; at least, we can be sure that
cyclic and ekpyrotic scenarios are possible for modified Finsler theories
\cite{stavrv} and we have a general geometric method for generating off--diagonal cosmological solutions in Finsler gravity. The properties of such solutions depend on the type of matter sources, symmetries and boundary/initial conditions we impose for our models. For stochastic off--diagonal components of metrics \cite%
{vstoch}, with momentum transfer, we reproduce the same problems for  overclosure  of universes as in low
energy string--brane limits with stochastic foam  \cite{mavromatos1,mavromatos2}. This can be considered as a  possible
direction for our further research.

Finally, it is worth mentioning an ''orthodox'' approach with Finsler like
and/or almost K\"{a}hler variables when the cosmological solutions are
derived for generic off--diagonal metrics in general relativity \cite%
{vqgrbr,vncbh}. Following this approach, we may conclude that there are not
modifications of Einstein gravity at classical level and that all
accelerating and anisotropic effects in our days cosmology are consequences
of certain nonlinear off--diagonal classical gravitational and matter field
interactions. Considering (in Finsler variables) nonholonomically deformed
FRW universes, we may model the bulk of dark energy and dark matter physics.
Such exact cosmological solutions can be constructed in explicit form \cite%
{vcosm1}. Nevertheless, we have to work with canonical Finsler gravity
models on (co) tangent bundles if quantum effects are taken into
consideration. Further developments will be provided in our papers under
elaboration. \vskip3pt

{\small \textbf{Acknowledgement: } I'm grateful to N. Mavromatos and P.
Stavrinos for important discussions, support and collaboration. The research
in this paper is partially supported by the Program IDEI,
PN-II-ID-PCE-2011-3-0256. Some recent results were communicated at the
parallel section "Quantum Gravity - Phenomenology", head G. Amelino-Camelia,
at Marcell Grossman 13, MG-13, Stockholm, Sweden (July 1-7, 2012). I thank
also the referee for very important critics and suggested new references
which I consider improved substantially the physical content of this paper.}

\appendix

\setcounter{equation}{0} \renewcommand{\theequation}
{A.\arabic{equation}} \setcounter{subsection}{0}
\renewcommand{\thesubsection}
{A.\arabic{subsection}}

\section{Formulas for N--adapted Coefficients}

For convenience, we present some important formulas in Finsler geometry and
applications in modern gravity, see details in \cite{vncg,ijgmmp,vrflg,vsgg}%
. Introducing the respective h-- v--components of the d--connection 1--form
for d--connection {\small $\mathbf{D}=\{\mathbf{\Gamma }_{\ \beta \gamma
}^{\alpha }\},$ we get the N--adapted coefficients \newline
$\mathbf{T}_{\ \beta \gamma }^{\alpha }=\{T_{\ jk}^{i},T_{\ ja}^{i},T_{\
ji}^{a},T_{\ bi}^{a},T_{\ bc}^{a}\}$ of torsion d--tensor (\ref{tors}),
\begin{eqnarray}
T_{\ jk}^{i} &=&L_{\ jk}^{i}-L_{\ kj}^{i},\ T_{\ ja}^{i}=-T_{\ aj}^{i}=C_{\
ja}^{i},\ T_{\ ji}^{a}=\Omega _{\ ji}^{a},\   \notag \\
T_{\ bi}^{a} &=&-T_{\ ib}^{a}=\frac{\partial N_{i}^{a}}{\partial y^{b}}-L_{\
bi}^{a},\ T_{\ bc}^{a}=C_{\ bc}^{a}-C_{\ cb}^{a}.  \label{dtors}
\end{eqnarray}
} A N--adapted differential form calculus allows us to derive the formulas
for h--v--components of curvature d--tensor (\ref{curv}) of a d--connection $%
\mathbf{D,}$ i.e. d--curvature $\mathbf{\mathbf{R}}_{\ \ \beta \gamma \delta
}^{\alpha }\mathbf{=}\{R_{\ hjk}^{i},R_{\ bjk}^{a},R_{\ jka}^{i},R_{\
bka}^{c},R_{\ jbc}^{i},R_{\ bcd}^{a}\},$ when {\small
\begin{eqnarray}
&&R_{\ hjk}^{i}=e_{k}L_{\ hj}^{i}-e_{j}L_{\ hk}^{i}+L_{\ hj}^{m}L_{\
mk}^{i}-L_{\ hk}^{m}L_{\ mj}^{i}-C_{\ ha}^{i}\Omega _{\ kj}^{a},
\label{dcurv} \\
&&R_{\ bjk}^{a}=e_{k}L_{\ bj}^{a}-e_{j}L_{\ bk}^{a}+L_{\ bj}^{c}L_{\
ck}^{a}-L_{\ bk}^{c}L_{\ cj}^{a}-C_{\ bc}^{a}\Omega _{\ kj}^{c},  \notag \\
&&R_{\ jka}^{i}=e_{a}L_{\ jk}^{i}-D_{k}C_{\ ja}^{i}+C_{\ jb}^{i}T_{\
ka}^{b},\ R_{\ bka}^{c}=e_{a}L_{\ bk}^{c}-D_{k}C_{\ ba}^{c}+C_{\ bd}^{c}T_{\
ka}^{c},  \notag \\
&&R_{\ jbc}^{i}=e_{c}C_{\ jb}^{i}-e_{b}C_{\ jc}^{i}+C_{\ jb}^{h}C_{\
hc}^{i}-C_{\ jc}^{h}C_{\ hb}^{i},  \notag \\
&&R_{\ bcd}^{a}=e_{d}C_{\ bc}^{a}-e_{c}C_{\ bd}^{a}+C_{\ bc}^{e}C_{\
ed}^{a}-C_{\ bd}^{e}C_{\ ec}^{a}.  \notag
\end{eqnarray}
} The values (\ref{dtors}) and (\ref{dcurv}) can be computed in explicit
form for the canonical d--connection $\widehat{\mathbf{\Gamma }}_{\ \alpha
\beta }^{\gamma }=\left( \widehat{L}_{jk}^{i},\widehat{L}_{bk}^{a},\widehat{C%
}_{jc}^{i},\widehat{C}_{bc}^{a}\right) $ with {\small
\begin{eqnarray}
\widehat{L}_{jk}^{i} &=&\frac{1}{2}g^{ir}\left( \mathbf{e}_{k}g_{jr}+\mathbf{%
e}_{j}g_{kr}-\mathbf{e}_{r}g_{jk}\right),  \notag \\
\widehat{L}_{bk}^{a} &=& e_{b}(N_{k}^{a})+\frac{1}{2}h^{ac}\left(
e_{k}h_{bc}-h_{dc}\ e_{b}N_{k}^{d}-h_{db}\ e_{c}N_{k}^{d}\right) ,
\label{candcon} \\
\widehat{C}_{jc}^{i} &=&\frac{1}{2}g^{ik}e_{c}g_{jk},\ \widehat{C}_{bc}^{a}=%
\frac{1}{2}h^{ad}\left( e_{c}h_{bd}+e_{c}h_{cd}-e_{d}h_{bc}\right) .  \notag
\end{eqnarray}%
} For any d--metric $\mathbf{g}$ on a N--anholonomic manifold $\mathbf{V,}$ $%
\widehat{\mathbf{D}}=\{\widehat{\mathbf{\Gamma }}_{\ \alpha \beta }^{\gamma
}\}$ satisfies the condition $\widehat{\mathbf{D}}\mathbf{g=}0$ vanishing of
''pure'' horizontal and vertical torsion coefficients, i. e. $\widehat{T}_{\
jk}^{i}=0$ and $\widehat{T}_{\ bc}^{a}=0,$ see formulas (\ref{dtors}). We
emphsize that, in general, $\widehat{T}_{\ ja}^{i},\widehat{T}_{\ ji}^{a}$
and $\widehat{T}_{\ bi}^{a}$ are not zero, but such nontrivial components of
torsion are induced by coefficients of an off--diagonal metric $\mathbf{g}%
_{\alpha \beta }$ (\ref{fansatz}).

Any geometric construction for the canonical d--connection $\widehat{\mathbf{%
D}}=\{\widehat{\mathbf{\Gamma }}_{\ \alpha \beta }^{\gamma }\}$ can be
re--defined equivalently into a similar one with the Levi--Civita connection
$\nabla =\{\Gamma _{\ \alpha \beta }^{\gamma }\}$ following formulas
\begin{equation}
\ \Gamma _{\ \alpha \beta }^{\gamma }=\widehat{\mathbf{\Gamma }}_{\ \alpha
\beta }^{\gamma }+\widehat{\mathbf{\ Z}}_{\ \alpha \beta }^{\gamma },
\label{deflc}
\end{equation}%
where N--adapted coefficients $\ $of connections, $\Gamma _{\ \alpha \beta
}^{\gamma }$ and $\widehat{\mathbf{\Gamma }}_{\ \alpha \beta }^{\gamma },$
and the distortion tensor $\ \widehat{\mathbf{\ Z}}_{\ \alpha \beta
}^{\gamma }$ are determined in unique forms by the coefficients of a metric $%
\mathbf{g}_{\alpha \beta }.$ The N--adapted components of the distortion
tensor $\widehat{\mathbf{\ Z}}_{\ \alpha \beta }^{\gamma }=\{\
Z_{jk}^{a},Z_{bk}^{i},Z_{bk}^{a},Z_{kb}^{i},Z_{jk}^{i},\
Z_{jb}^{a},Z_{bc}^{a},Z_{ab}^{i}\}$ are {\small
\begin{eqnarray}
\ Z_{jk}^{a} &=&-\widehat{C}_{jb}^{i}g_{ik}h^{ab}-\frac{1}{2}\Omega
_{jk}^{a},~Z_{bk}^{i}=\frac{1}{2}\Omega _{jk}^{c}h_{cb}g^{ji}-\Xi _{jk}^{ih}~%
\widehat{C}_{hb}^{j},  \notag \\
Z_{bk}^{a} &=&~^{+}\Xi _{cd}^{ab}~\widehat{T}_{kb}^{c},\ Z_{kb}^{i}=\frac{1}{%
2}\Omega _{jk}^{a}h_{cb}g^{ji}+\Xi _{jk}^{ih}~\widehat{C}_{hb}^{j},\
Z_{jk}^{i}=0,  \label{deft} \\
\ Z_{jb}^{a} &=&-~^{-}\Xi _{cb}^{ad}~\widehat{T}_{jd}^{c},\ Z_{bc}^{a}=0,\
Z_{ab}^{i}=-\frac{g^{ij}}{2}\left[ \widehat{T}_{ja}^{c}h_{cb}+\widehat{T}%
_{jb}^{c}h_{ca}\right] ,  \notag
\end{eqnarray}%
for $\ \Xi _{jk}^{ih}=\frac{1}{2}(\delta _{j}^{i}\delta
_{k}^{h}-g_{jk}g^{ih}),~^{\pm }\Xi _{cd}^{ab}=\frac{1}{2}(\delta
_{c}^{a}\delta _{d}^{b}+h_{cd}h^{ab})$ and $\widehat{T}_{\ ja}^{c}=\widehat{L%
}_{aj}^{c}-e_{a}(N_{j}^{c}).\ $ }

\setcounter{equation}{0} \renewcommand{\theequation}
{B.\arabic{equation}} \setcounter{subsection}{0}
\renewcommand{\thesubsection}
{B.\arabic{subsection}}

\section{A Class of General Cosmological Solutions}

\label{asect2}

The Einstein equations for $\widehat{\mathbf{D}}$ computed for ansatz (\ref%
{ansgensol}) and source {\small
\begin{eqnarray}
\Upsilon _{\ \ \quad ^{3}\beta }^{\ ^{3}\alpha } &=&diag[\Upsilon _{\
^{3}\gamma };\Upsilon _{1}=\Upsilon _{2}=\Upsilon _{2}(x^{k});\Upsilon
_{3}=\Upsilon _{4}=\Upsilon _{4}(x^{k},\ ^{0}v);  \notag \\
\Upsilon _{5} &=&\Upsilon _{6}=\Upsilon _{6}(u^{\ ^{0}\alpha },\
^{1}v);\Upsilon _{7}=\Upsilon _{8}=\Upsilon _{8}(u^{\ ^{1}\alpha },\ ^{2}v)]
\label{s8d}
\end{eqnarray}
when the partial derivatives, for instance, are parameterized $\partial _{\
^{1}v}=\partial /\partial \ ^{1}v=$ $\partial /\partial y^{5},\ $ $\partial
_{\ ^{2}v}=\partial /\partial \ ^{2}v=$ $\partial /\partial y^{7},$ and $%
N_{\ ^{0}\alpha }^{5}=\ ^{1}w_{\ ^{0}\alpha }(u^{\ ^{0}\alpha },\ ^{1}v),\
N_{\ ^{0}\alpha }^{6}=\ ^{1}n_{\ ^{0}\alpha }(u^{\ ^{0}\alpha },\ ^{1}v),$ $%
N_{\ ^{1}\alpha }^{7}=\ ^{2}w_{\ ^{1}\alpha }(u^{\ ^{1}\alpha },\ ^{2}v),\
N_{\ ^{1}\alpha }^{8}=\ ^{2}n_{\ ^{1}\alpha }(u^{\ ^{1}\alpha },\ ^{2}v).$ }
For zero N--connection coefficients $N_{i}^{a},$ with $i,j,...=1,2,3,4$ and $%
a,b,..=5,6,7,8,$ we can chose such solutions for $h_{a}$ when (\ref{8dods})
have certain limits to the diagonal cosmological metric (\ref{diagfans}).
Such a very general off--diagonal, inhomogeneous and locally anisotropic
cosmological dynamics, with one Killing symmetry vector $\partial /\partial
y^{8}=$ $\partial /\partial \ ^{v}\varphi $ (a similar class of solutions
can be generated if as $y^{8}$ we take $\ ^{v}\theta $ for $y^{7}=\
^{v}\varphi )$ is determined by coefficients {\small
\begin{eqnarray*}
&&h_{5} =\ _{1}^{0}h(x^{1},\ ^{h}r,t,\ ^{h}\theta )\ [\partial _{\ ^{v}r}\
^{1}f(x^{1},\ ^{h}r,t,\ ^{h}\theta ,\ ^{v}r)]^{2} |\ ^{1}\varsigma (x^{1},\
^{h}r,t,\ ^{h}\theta ,\ \ ^{v}r)|, \\
&&h_{6} =[\ ^{1}f(x^{1},\ ^{h}r,t,\ ^{h}\theta ,\ \ \ ^{v}r)-\
_{1}^{0}f(x^{1},\ ^{h}r,t,\ ^{h}\theta )]^{2}; \\
&&\eta _{5} =[f(\ ^{h}r,t,\ ^{h}\theta )-\ ^{0}f(\ ^{h}r,t)]^{2},\check{h}%
_{5}=\ ^{h}a^{2}(t)\ (\ ^{h}r)^{2}\sin ^{2}\ ^{h}\theta \\
&&w_{\ ^{0}\beta } =-\partial _{\ ^{0}\beta }\ ^{1}\varsigma (x^{1},\
^{h}r,t,\ ^{h}\theta , \ ^{v}r)/\partial _{\ \ ^{v}r}\ ^{1}\varsigma
(x^{1},\ ^{h}r,t,\ ^{h}\theta ,\ \ ^{v}r), \\
&&n_{\ ^{0}\beta } =\ ^{1}n_{\ ^{0}\beta }(x^{1},\ ^{h}r,t,\ ^{h}\theta )+\
^{2}n_{\ ^{0}\beta}(x^{1},\ ^{h}r,t,\ ^{h}\theta)\int d\ ^{v}r \
^{1}\varsigma (x^{1},\ ^{h}r,t,\ ^{h}\theta,\ ^{v}r) \\
&&\ [\partial _{\ ^{v}r}\ ^{1}f(x^{1},\ ^{h}r,t,\ ^{h}\theta ,\ ^{v}r]^{2}
\lbrack \ ^{1}f(x^{1},\ ^{h}r,t,\ ^{h}\theta ,\ ^{v}r)-\ _{1}^{0}f(x^{1},\
^{h}r,t,\ ^{h}\theta )]^{-3},
\end{eqnarray*}
for $\ ^{1}\varsigma =\ _{1}^{0}\varsigma (x^{1},\ ^{h}r,t,\ ^{h}\theta )-%
\frac{\ ^{1}\Upsilon _{2}}{8}\ _{1}^{0}h(x^{1},\ ^{h}r,t,\ ^{h}\theta ) \int
d\ \ ^{v}r\ \lbrack \partial _{\ ^{1}v}\ ^{1}f(x^{1},\ ^{h}r,t,\ ^{h}\theta
,\ \ ^{v}r)]$ \newline
$[\ ^{1}f(x^{1},\ ^{h}r,t,\ ^{h}\theta ,\ \ ^{v}r)-\ _{1}^{0}f(x^{1},\
^{h}r,t,\ ^{h}\theta )];$ for any generation $\ ^{1}f(x^{1},\ ^{h}r,t,\
^{h}\theta ,\ ^{v}r)]^{2}$ and integration functions $\ _{1}^{0}h(x^{1},\
^{h}r,t,\ ^{h}\theta )$ and small parameter $\varepsilon $, $\
_{1}^{0}h(x^{1},\ ^{h}r,t,\ ^{h}\theta )$\newline
$[\partial _{\ ^{1}v}\ ^{1}f(x^{1},\ ^{h}r,t,\ ^{h}\theta ,\ \ ^{v}r)]^{2}\
|\ ^{1}\varsigma (x^{1},\ ^{h}r,t,\ ^{h}\theta ,\ ^{v}r)| =\ ^{v}a^{2}(t)
(1+\varepsilon \xi (x^{1},\ ^{h}r,t,\ ^{h}\theta ,\ ^{v}r))$, and $w_{\
^{1}\beta }=0,n_{\ ^{1}\beta }=0,h_{7}=\ ^{v}a^{2}(t)(\ ^{v}r)^{2},h_{8}=\
^{v}a^{2}(t)\ (\ ^{v}r)^{2}\sin ^{2}\ ^{v}\theta .$}\footnote{%
We considered, for example, only small contributions from the velocity $h$%
--subspace to the ''real'' $v$--subspaces.}

For simplicity, we can fix $\ ^{1}n_{\ ^{0}\beta }=0$ and that $\
^{1}f(x^{1},\ ^{h}r,t,\ ^{h}\theta ,\ \ ^{v}r)$ induces $\xi =\xi (t,\
^{h}\theta ,\ ^{v}r)$ as a solution of any three dimensional solitonic
(nonlinear wave) equation, for instance, of type $\frac{\partial ^{2}\xi }{%
\partial (\ ^{v}r)^{2}}+\epsilon (\xi ^{\prime }+6\xi \ \xi ^{\ast }+\xi
^{\ast \ast \ast })^{\ast }=0,\ \epsilon =\pm 1,$ where $\xi ^{\prime
}=\partial \xi /\partial t$ and $\xi ^{\ast }=\partial \xi /\partial \
^{h}\theta .$ Such solitons are stable and generate solitonic configurations
for the metric and nontrivial N--connection coefficients (for $\ ^{0}\beta
=3,4),$ $\widetilde{\eta }_{5}=\widetilde{\eta }_{5}[\xi ]\sim 1+\varepsilon
\varpi _{5}[\xi ],\widetilde{\eta }_{6}=\widetilde{\eta }_{6}[\xi ]\sim
1+\varepsilon \varpi _{6}[\xi ],$ and $\widetilde{w}_{3}\rightarrow
\varepsilon \widetilde{w}_{3}[\xi ],\widetilde{w}_{4}\rightarrow \varepsilon
\widetilde{w}_{4}[\xi ],\widetilde{n}_{3}\rightarrow \varepsilon \widetilde{n%
}_{3}[\xi ],\widetilde{n}_{4}\rightarrow \varepsilon \widetilde{n}_{4}[\xi ],
$ where, for simplicity, we fixed the boundary conditions to have a
functional dependence on $\xi $ and vanishing values if $\varepsilon
\rightarrow 0.$

\end{document}